\documentclass[ALICE,manyauthors]{cernphprep}
\usepackage[comma,square,numbers,sort&compress]{natbib}
\usepackage{hyperref}
\usepackage{lineno}
\usepackage{xspace}
\usepackage{multirow}
\usepackage{placeins}
\usepackage{amsmath}
\usepackage{breqn}
\usepackage[T1]{fontenc}

\DeclareMathOperator\erf{erf}
\begin{document}
%%%%%%%%%%%%%%%%%%%%%%%%%%%%%%%%%%%%%%%%%%%%%%%%%%
% These are some new commands that may be useful 
% for paper writing in general. If other newcommands
% are needed for your specific paper, please feel 
% free to add here. 
%
% The currently available commands are organized in: 
% 1) Systems
% 2) Quantities
% 3) Energies and units
% 4) Detectors
% 5) particle species 
%%%%%%%%%%%%%%%%%%%%%%%%%%%%%%%%%%%%%%%%%%%%%%%%%%

\mathchardef\mhyphen="2D

% 1) SYSTEMS 
\newcommand{\pp}           {pp\xspace}
\newcommand{\ppbar}        {\mbox{$\mathrm {p\overline{p}}$}\xspace}
\newcommand{\XeXe}         {\mbox{Xe--Xe}\xspace}
\newcommand{\PbPb}         {\mbox{Pb--Pb}\xspace}
\newcommand{\pA}           {\mbox{pA}\xspace}
\newcommand{\pPb}          {\mbox{p--Pb}\xspace}
\newcommand{\AuAu}         {\mbox{Au--Au}\xspace}
\newcommand{\dAu}          {\mbox{d--Au}\xspace}

% 2) QUANTITIES 
\newcommand{\sigmapid}{$\sigma^{^{3}\mathrm{He}}_{\mathrm{d}E/\mathrm{d}x}$}
\newcommand{\nsigma} {$\left(  \mathrm{d}E/\mathrm{d}x -  \langle \mathrm{d}E/\mathrm{d}x \rangle_{^{3}\mathrm{He}} \right) /\sigma^{^{3}\mathrm{He}}_{\mathrm{d}E/\mathrm{d}x}$ }
\newcommand{\s}            {\ensuremath{\sqrt{s}}\xspace}
\newcommand{\snn}          {\ensuremath{\sqrt{s_{\mathrm{NN}}}}\xspace}
\newcommand{\pt}           {\ensuremath{p_{\rm T}}\xspace}
\newcommand{\mt}           {\ensuremath{m_{\rm T}}\xspace}
\newcommand{\meanpt}       {$\langle p_{\mathrm{T}}\rangle$\xspace}
\newcommand{\ycms}         {\ensuremath{y_{\rm CMS}}\xspace}
\newcommand{\ylab}         {\ensuremath{y_{\rm lab}}\xspace}
\newcommand{\etarange}[1]  {\mbox{$\left | \eta \right |~<~#1$}}
\newcommand{\yrange}[1]    {\mbox{$\left | y \right |~<~#1$}}
\newcommand{\dndy}         {\ensuremath{\mathrm{d}N_{\mathrm{ch}}/\mathrm{d}y}\xspace}
\newcommand{\dndeta}       {\ensuremath{\mathrm{d}N_{\mathrm{ch}}/\mathrm{d}\eta}\xspace}
\newcommand{\avdndeta}     {\ensuremath{\langle\dndeta\rangle}\xspace}
\newcommand{\dNdy}         {\ensuremath{\mathrm{d}N_{\mathrm{ch}}/\mathrm{d}y}\xspace}
\newcommand{\Npart}        {\ensuremath{N_\mathrm{part}}\xspace}
\newcommand{\Ncoll}        {\ensuremath{N_\mathrm{coll}}\xspace}
\newcommand{\dEdx}         {\ensuremath{\textrm{d}E/\textrm{d}x}\xspace}
\newcommand{\RpPb}         {\ensuremath{R_{\rm pPb}}\xspace}

% 3) ENERGIES, UNITS
\newcommand{\nineH}        {$\sqrt{s}~=~0.9$~Te\kern-.1emV\xspace}
\newcommand{\seven}        {$\sqrt{s}~=~7$~Te\kern-.1emV\xspace}
\newcommand{\twoH}         {$\sqrt{s}~=~0.2$~Te\kern-.1emV\xspace}
\newcommand{\twosevensix}  {$\sqrt{s}~=~2.76$~Te\kern-.1emV\xspace}
\newcommand{\five}         {$\sqrt{s}~=~5.02$~Te\kern-.1emV\xspace}
\newcommand{\twosevensixnn}{$\sqrt{s_{\mathrm{NN}}}~=~2.76$~Te\kern-.1emV\xspace}
\newcommand{\fivenn}       {$\sqrt{s_{\mathrm{NN}}}~=~5.02$~Te\kern-.1emV\xspace}
\newcommand{\LT}           {L{\'e}vy-Tsallis\xspace}
\newcommand{\GeVc}         {Ge\kern-.1emV/$c$\xspace}
\newcommand{\MeVc}         {Me\kern-.1emV/$c$\xspace}
\newcommand{\TeV}          {Te\kern-.1emV\xspace}
\newcommand{\GeV}          {Ge\kern-.1emV\xspace}
\newcommand{\MeV}          {Me\kern-.1emV\xspace}
\newcommand{\GeVmass}      {Ge\kern-.2emV/$c^2$\xspace}
\newcommand{\MeVmass}      {Me\kern-.2emV/$c^2$\xspace}
\newcommand{\lumi}         {\ensuremath{\mathcal{L}}\xspace}

% 4) DETECTORS 
\newcommand{\ITS}          {\rm{ITS}\xspace}
\newcommand{\TOF}          {\rm{TOF}\xspace}
\newcommand{\ZDC}          {\rm{ZDC}\xspace}
\newcommand{\ZDCs}         {\rm{ZDCs}\xspace}
\newcommand{\ZNA}          {\rm{ZNA}\xspace}
\newcommand{\ZNC}          {\rm{ZNC}\xspace}
\newcommand{\SPD}          {\rm{SPD}\xspace}
\newcommand{\SDD}          {\rm{SDD}\xspace}
\newcommand{\SSD}          {\rm{SSD}\xspace}
\newcommand{\TPC}          {\rm{TPC}\xspace}
\newcommand{\TRD}          {\rm{TRD}\xspace}
\newcommand{\VZERO}        {\rm{V0}\xspace}
\newcommand{\VZEROA}       {\rm{V0A}\xspace}
\newcommand{\VZEROC}       {\rm{V0C}\xspace}
\newcommand{\Vdecay} 	   {\ensuremath{V^{0}}\xspace}

% 4) PARTICLE SPECIES 
\newcommand{\ee}           {\ensuremath{e^{+}e^{-}}} 
\newcommand{\pip}          {\ensuremath{\pi^{+}}\xspace}
\newcommand{\pim}          {\ensuremath{\pi^{-}}\xspace}
\newcommand{\kap}          {\ensuremath{\rm{K}^{+}}\xspace}
\newcommand{\kam}          {\ensuremath{\rm{K}^{-}}\xspace}
\newcommand{\pbar}         {\ensuremath{\rm\overline{p}}\xspace}
\newcommand{\kzero}        {\ensuremath{{\rm K}^{0}_{\rm{S}}}\xspace}
\newcommand{\lmb}          {\ensuremath{\Lambda}\xspace}
\newcommand{\almb}         {\ensuremath{\overline{\Lambda}}\xspace}
\newcommand{\Om}           {\ensuremath{\Omega^-}\xspace}
\newcommand{\Mo}           {\ensuremath{\overline{\Omega}^+}\xspace}
\newcommand{\X}            {\ensuremath{\Xi^-}\xspace}
\newcommand{\Ix}           {\ensuremath{\overline{\Xi}^+}\xspace}
\newcommand{\Xis}          {\ensuremath{\Xi^{\pm}}\xspace}
\newcommand{\Oms}          {\ensuremath{\Omega^{\pm}}\xspace}
\newcommand{\degree}       {\ensuremath{^{\rm o}}\xspace}

%%%%%%%%%%%%%%%%%%%%%  Title page %%%%%%%%%%%%%%%%%%%%%%%%%%%%%%
\begin{titlepage}
% the dates below correspond to CERN approval
% please don't touch: EB chairs will take care
\PHyear{2021}       % required, will be obtained from CERN
\PHnumber{194}      % required, will be obtained from CERN
\PHdate{24 September}  % required, will be obtained from CERN
%%%%%%%%%%%%%%%%%%%%%%%%%%%%%%%%%%%%%%%%%%%%%%%%%%%%%%%%%%%%%%%%

\title{Production of light (anti)nuclei in pp collisions at $\sqrt{s} = $ 13~TeV}
\ShortTitle{(Anti)nuclei in pp collisions at $\sqrt{s} = $ 13~TeV}

\Collaboration{ALICE Collaboration\thanks{See Appendix~\ref{app:collab} for the list of collaboration members}}
\ShortAuthor{ALICE Collaboration} 

\begin{abstract}
Understanding the production mechanism of light (anti)nuclei is one of the key challenges of nuclear physics and has important consequences for astrophysics, since it provides an input for indirect dark-matter searches in space. In this paper, the latest results about the production of light (anti)nuclei in pp collisions at \s $= 13$~TeV are presented, focusing on the comparison with the predictions of coalescence and thermal models. For the first time, the coalescence parameters $B_2$ for deuterons and $B_3$ for helions are compared with parameter-free theoretical predictions that are directly constrained by the femtoscopic measurement of the source radius in the same event class. A fair description of the data with a Gaussian wave function is observed for both deuteron and helion, supporting the coalescence mechanism for the production of light (anti)nuclei in pp collisions. This method paves the way for future investigations of the internal structure of more complex nuclear clusters, including the hypertriton.

\end{abstract}
\end{titlepage}

\setcounter{page}{2} %please do not remove this line

\section{Introduction} 
\label{sec:Introduction}

In high-energy hadronic collisions at the LHC, the production of light (anti)nuclei and more complex multi-baryon bound states, such as (anti)hypertriton~\cite{hypertriton_PbPb_276}, is observed. An unexpectedly large yield of light nuclei was observed for the first time in proton-nucleus collisions at the CERN PS accelerator~\cite{NucleiCERNPS}. Twenty-five years later, the study of nuclear production was carried out at Brookhaven AGS and at CERN SPS, with the beginning of the program of relativistic nuclear collisions~\cite{ReviewAGS}. Extensive studies of the production of light (anti)nuclei were later performed at the Relativistic Heavy-Ion Collider (RHIC)~\cite{RHIC1,RHIC2,RHIC3,RHIC4}, including the first observation of an antihypernucleus~\cite{RHIC5} and of anti(alpha)~\cite{RHIC6}. In this paper, the focus will be on results obtained at LHC. The production yields of light (anti)nuclei have been measured as a function of transverse momentum (\pt) and charged-particle multiplicity in different collision systems and at different center-of-mass energies by ALICE~\cite{nuclei_pp_PbPb,deuteron_pp_7TeV,deuteron_PbPb_276TeV,nuclei_pp,4He_PbPb,deuteron_pPbALICE,3He_pPb,deuteron_pp_13TeV}. One of the most interesting observations obtained from such a large variety of experimental data is that the production of light (anti)nuclei seems to depend solely on the charged-particle multiplicity (hereinafter denoted multiplicity). This observation manifests itself in the continuous evolution of the deuteron-to-proton (d/p) and $^{3}$He-to-proton ($^{3}$He/p) ratios with the event multiplicity across different collision systems and energies~\cite{3He_pPb, deuteron_pp_13TeV}. The results presented in this paper complement the existing picture, providing measurements in yet unexplored multiplicity regions.

These measurements have an important astrophysical value as they provide input for the background estimates in indirect dark matter searches in space. Indeed, only small systems like pp collisions are relevant for such searches because the interstellar medium consists mostly of hydrogen (protons) and helium (alpha particles). In this context, the observation of a significant antimatter excess with respect to the expected background of antimatter produced in ordinary cosmic ray pp or p--alpha interactions would represent a signal for dark matter annihilation in the galactic halo or for the existence of antimatter islands in our universe~\cite{dark_matter1,dark_matter2,dark_matter3,dark_matter4}. 

The theoretical description of the production mechanism of (anti)nuclei is still an open problem and under intense debate in the scientific community.
Two phenomenological models are typically used to describe the production of multi-baryon bound states: the statistical hadronisation model (SHM)~\cite{SHM1,SHM2,SHM3,SHM4,SHM5,SHM6,vanillaCSM} and the coalescence model~\cite{Coalescence1,Coalescence2,iEBE_VISHNU,Coalescence3,coalescenceSmallSystems,alternativeCoalescence}. In the former, light nuclei are assumed to be emitted by a source in local thermal and hadrochemical equilibrium and their abundances are fixed at chemical freeze-out. The version of this model using the grand-canonical ensemble reproduces the light-flavoured hadron yields measured in central nucleus--nucleus collisions, including those of (anti)nuclei and (anti)hypernuclei~\cite{SHM1}. In pp and p--Pb collisions, the production of light nuclei can be described using a different implementation of this model based on the canonical ensemble, where exact conservation of the electric charge, strangeness, and baryon quantum numbers is applied within a pre-defined correlation volume~\cite{SHM4,vanillaCSM}. In the coalescence model, light nuclei are assumed to be formed by the coalescence of protons and neutrons which are close in phase space at kinetic freeze-out~\cite{Coalescence2}. In the most simple version of this model, nucleons are treated as point-like particles and only correlations in momentum space are considered, i.e. the bound state is assumed to be formed if the difference between the momenta of nucleons is smaller than a given threshold $p_{0}$, a free parameter of the model which is typically of the order of 100 MeV$/\textit{c}$. This simple version of the coalescence model can approximately reproduce deuteron production data in low-multiplicity collisions and was recently used to describe the jet-associated deuteron $p_{\rm T}$-differential yields in pp collisions at $\sqrt{s} = $13~TeV~\cite{DeuteronInJets}.
In recent developments~\cite{iEBE_VISHNU,coalescence_correlations}, the quantum-mechanical properties of nucleons and nuclei are taken into account and the coalescence probability is calculated from the overlap between the source function of the emitted protons and neutrons, which are mapped on the Wigner density of the nucleus. This state-of-the-art coalescence model describes the d/p and $^{3}$He/p ratios measured in different collision systems as a function of multiplicity~\cite{coalescenceSmallSystems}. On the contrary, the simple coalescence approach provides a description of \pt spectra of light (anti)nuclei measured in high-energy hadronic collisions only in the low-multiplicity regime~\cite{deuteron_pPbALICE}. 

In this paper, the measurement of the production yields of light (anti)nuclei in pp collisions at $\sqrt{s} = $13~TeV are presented. In particular, part of the results is obtained from data collected with a high-multiplicity trigger (see Sec.~\ref{sec:ExperimentalApparatus}), accessing a multiplicity typically obtained in p--Pb and peripheral Pb--Pb collisions. For the first time, the yields of (anti)nuclei are measured in a multiplicity region in which high-precision femtostopic measurements of the source size~\cite{sourceSizeHMpp} are available. This allows for a parameter-free comparison of the coalescence measurements with theoretical calculations, showing the potential of this technique to set constraints on the wave function of (anti)nuclei.

\section{Detector and data sample} 
\label{sec:ExperimentalApparatus}

%ALICE is a general-purpose experiment at the LHC dedicated to the study of heavy-ion and hadronic collisions at ultra-relativistic energies.
A detailed description of the ALICE apparatus and its performance can be found in Refs.~\cite{ALICEexperiment} and~\cite{ALICEperformance}. 
The trajectories of charged particles are reconstructed in the ALICE central barrel with the Inner Tracking System (ITS)~\cite{ITS} and the Time Projection Chamber (TPC)~\cite{TPC}. The ITS consists of six cylindrical layers of silicon detectors and the two innermost layers form the Silicon Pixel Detector (SPD). The ITS is used for the reconstruction of primary and secondary vertices and of charged-particle trajectories. The TPC is used for track reconstruction, charged-particle momentum measurement and for charged-particle identification via the measurement of their specific energy loss (\dEdx) in the TPC gas~\cite{ALICEperformance}.
Particle identification at high momentum is complemented by the time-of-flight measurement provided by the TOF detector~\cite{TOF}. The aforementioned detectors are located inside a large solenoid magnet, which provides a homogeneous magnetic field of 0.5~T parallel to the beam line, and cover the pseudorapidity interval $|\eta| < 0.9$.
Collision events are triggered by two plastic scintillator arrays, V0C and V0A~\cite{VZEROPerformance}, located along the beam axis of the interaction point, covering the pseudorapidity regions \mbox{$-$3.7 $< \eta <$ $-$1.7} and \mbox{2.8 $< \eta <$ 5.1}, respectively. The signals from V0A and V0C are summed to form the V0M signal, which is used to define event classes to which the measured multiplicity is associated~\cite{mutliplicity7tev}. Moreover, the timing information of the V0 detectors is used for the offline rejection of events triggered by interactions of the beam with the residual gas in the LHC vacuum pipe.

The results presented in this paper are obtained from data collected in 2016, 2017 and 2018, both with minimum bias (MB) and high multiplicity (HM) triggers. For the minimum-bias event trigger, coincident signals in both V0 scintillators are required to be synchronous with the beam crossing time defined by the LHC clock. Events with high charged-particle multiplicities are triggered on by additionally requiring the total signal amplitude measured in the V0 detector to exceed a threshold. At the analysis level, the 0--0.1\% percentile of inelastic events with the highest V0 multiplicity (V0M) is selected to define the high-multiplicity event class. Events with multiple vertices identified with the SPD are tagged as pile-up in the same bunch crossing (in-bunch pile-up) and removed from the analysis~\cite{ALICEperformance}. Assuming that all the in-buch pile-up is in the 0--0.01\% percentile of inelastic events, which is the worst-case scenario, only 3\% of the selected events (in the 0--0.01\% percentile) would be pile-up events. Therefore, the effect of in-bunch pile-up on the production spectra is negligible. Pile-up in different bunch crossings, instead, is rejected by requiring track hits in the SPD and its contribution is negligible. The data sample consists of approximately 2.6 billion MB events and 650 million HM events.

For the measurements of (anti)protons and (anti)deuterons, the high-multiplicity data sample is divided into three multiplicity classes: HM I, HM II and HM III. The multiplicity classes are determined from the sum of the signal amplitudes measured by the V0 detectors and defined in terms of the percentiles of the \mbox{INEL $>$ 0} pp cross section, where an \mbox{INEL $>$ 0} event is a collision with at least a charged particle in the pseudorapidity region $|\eta| < 1$~\cite{strangeness_vs_mult}. For this purpose, charged particles are measured with SPD tracklets, obtained from a pair of hits in the first and second layer of the SPD, respectively. In the case of (anti)triton (${}^{3}\mathrm{H}$) and (anti)helion (${}^{3}\mathrm{He}$), due to their lower production rate, it is not possible to divide the HM sample into smaller classes, but for (anti)helion the MB sample is divided into two multiplicity classes, MB I and MB II, defined from the percentiles of the \mbox{INEL $>$ 0} pp cross section. The average charged particle multiplicity \avdndeta for all the multiplicity classes will be reported in Tab.~\ref{tab:multiplicity}. It is defined as the number of primary charged particles produced in the pseudorapidity interval $|\eta| < 0.5$. A detailed description of the \avdndeta estimation can be found in Ref.~\cite{multiplicity_measurement}.

\section{Data analysis}

\subsection{Track selection}
\label{sec:track_selection}

(Anti)nuclei candidates are selected from the charged-particle tracks reconstructed in the ITS and TPC in the pseudorapidity interval $|\eta| < 0.8$. The criteria used for track selection are the same as reported in Ref.~\cite{deuteron_pp_13TeV}. Particle identification is performed using the \dEdx measured by the TPC and the time-of-flight measured by the TOF. For the TPC analysis, the signal is obtained from the $n\sigma_{\mathrm{TPC}}$ distribution, where $n\sigma_{\mathrm{TPC}}$ is the difference between the measured and expected signals for a given particle hypothesis, divided by the resolution. For the TOF analysis, the yield in each \pt interval is extracted from the distribution of the TOF squared-mass, defined as $m^{2} = p^{2}\left(t_{\mathrm{TOF}}^2/L^2 - 1/c^2\right)$, where $t_{\mathrm{TOF}}$ is the measured time-of-flight, $L$ is the length of the track and $p$ is the momentum of the particle. Similarly to the TPC case, one defines $n\sigma_{\mathrm{TOF}}$ as the difference between the measured and expected time of flight for a given particle hypothesis, divided by the resolution. For the TOF analysis, a pre-selection based on the measured TPC \dEdx ($|n\sigma_{\mathrm{TPC}}|<3$) is performed to reduce the background originating from other particle species. More details about particle identification with TPC and with TOF can be found in Ref.~\cite{deuteron_pp_13TeV}.

The (anti)deuteron yield is extracted from the TPC signal for $p_{\mathrm{T}} < $ 1 GeV$/$\textit{c}, while at higher \pt the yield is extracted from the TOF after the pre-selection using the TPC signal. For (anti)protons, the TOF is used for the entire \pt range. (Anti)tritons are identified through the TPC signal and after a pre-selection with TOF ($|n\sigma_{\mathrm{TOF}}|<3$) for $p_{\mathrm{T}} < $ 2 GeV$/$\textit{c}.
The (anti)helion identification is based only on the TPC \dEdx, which provides a good separation of its signal from that of other particle species. This is due to the charge $\mathrm{Z}=2$ of this nucleus.

\subsection{Efficiency and acceptance correction}
\label{sec:efficiency}

The estimation of reconstruction efficiencies of both nuclei and antinuclei, as well as those of the contamination to the raw \pt spectra of nuclei from spallation and of the signal loss due to event selection, requires Monte Carlo (MC) simulations.
Simulated pp collision events, generated using Pythia 8~\cite{Pythia8} (Monash 2013 tune~\cite{Pythia8Monash2013}), are enriched by an injected sample of (anti-)nuclei generated with a flat \pt distribution in the transverse-momentum range $0 < p_{\mathrm{T}} < 10\ \mathrm{GeV}/\textit{c}$ and a flat rapidity distribution in the range $|y| < 1$. The interactions of the generated particles with the experimental apparatus are modeled by GEANT4~\cite{GEANT4}. The detector conditions during the data taking are reproduced in the simulations.

The raw \pt spectra of (anti)nuclei are corrected for the reconstruction efficiency and acceptance, defined as 

\begin{equation}
\epsilon (p_{\mathrm{T}}) = \frac{N_{\mathrm{rec}} (|\eta|<0.8, |y|<0.5, p^{\mathrm{rec}}_{\mathrm{T}})}{N_{\mathrm{gen}} (|y|<0.5, p_{\mathrm{T}})} ,
\end{equation}

where $N_{\mathrm{rec}}$ and $N_{\mathrm{gen}}$ are the number of reconstructed and generated (anti)nuclei, respectively. The same criteria for the track selection and particle identification used in the real data analysis are applied to reconstructed tracks in the simulation. Considering that (anti)nuclei are injected with a flat \pt distribution into the simulated events, their input distributions are reshaped using \pt-dependent weights to match the real shape observed in data. The latter is parameterised using a L\'evy-Tsallis function whose parameters are determined using an iterative procedure: they are initialized using the values taken from Ref.~\cite{deuteron_pp_13TeV}, the corrected spectrum is then fitted with the same function and a new set of parameters is determined and used for the next iteration. The parameters are found to converge after two iterations, with differences from the previous one of less than one per mille.  
Protons are abundantly produced by Pythia 8 and their spectral shape is consistent with the one obtained in real data. For this reason, (anti)protons are not injected additionally into the simulation and their shape is not modified. 

\subsection{Fraction of primary nuclei}
Secondary nuclei are produced in the interaction of particles with the detector material. To obtain the yields of primary nuclei produced in a collision, the number of secondaries must be subtracted from the measured yield. Since the production of secondary antinuclei is extremely rare, this correction is applied only to nuclei and not to antinuclei. For (anti)protons, instead, also a contribution from weak decays of heavier unstable particles (for example $\Lambda$ hyperons) is present and cannot be neglected. The fraction of primary nuclei is evaluated using different techniques according to the analysis.

For deuterons and (anti)protons, the primary fraction is obtained by fitting the distribution of the measured distance of closest approach to the primary vertex in the transverse plane (DCA$_{xy}$). For the fit, templates obtained from MC are used, as described in Ref.~\cite{nuclei_pp_PbPb}. The DCA$_{xy}$ distribution of anti-deuterons is used as a template for primary deuterons considering the negligible feed-down from weak decays of hypertriton. The production of secondary deuterons is more relevant at low $p_{\mathrm{T}}$ (at $p_{\mathrm{T}} = $~0.7~GeV$/$\textit{c} the fraction of secondary deuterons is $\sim$ 40\%), decreases exponentially with the transverse momentum ($<$ 5\% for $p_{\mathrm{T}} = $~1.4~GeV$/$\textit{c}) and becomes negligible for $p_{\mathrm{T}} > $~1.6~GeV$/$\textit{c}. For the (anti)proton analysis, all the templates are taken from MC. In this case, also a template for weak decay is used. The fraction of secondary protons from material is maximum at low transverse momentum ($\sim$5\% for $p_{\mathrm{T}} = $~0.4~GeV$/$\textit{c}) and decreases exponentially, becoming negligible for $p_{\mathrm{T}} > $ 1~GeV$/$\textit{c}. Also, the fraction of secondary (anti)protons from weak decays is maximum at low transverse momentum ($\sim$30\% for $p_{\mathrm{T}} = $~0.4~GeV$/$\textit{c}) and decreases exponentially ($\sim$10\% for $p_{\mathrm{T}} = $~5~GeV$/$\textit{c}).

For helion and triton, the primary fraction is obtained by fitting the DCA$_{xy}$ distribution with two Gaussian functions with different widths, one for the primary and one for the secondary nuclei, respectively. The spallation background is also fitted using a parabola and a constant function to estimate the systematic uncertainties. For the latter, variations of the binning and fit range are also considered. A smooth function can be used in this case, considering that the peak in the DCA$_{xy}$ distribution of secondary nuclei, which typically appears close to zero and is caused by the wrong association of one SPD cluster to the track during reconstruction, is negligible for $p_{\mathrm{T}} > 1.5$~GeV$/c$. The fraction of secondary helions and tritons, for both MB and HM pp collisions, are found to be about 15$\%$ in the \pt interval $1.5 < p_{\mathrm{T}} < 2$ GeV$/$\textit{c}, about 3$\%$ in the \pt interval $2 < p_{\mathrm{T}} < 2.5$ GeV$/$\textit{c} and negligible for higher \pt. For tritons and helions, only antinuclei are used for $p_{\mathrm{T}} < 1.5$~GeV$/$\textit{c}, where the secondary fraction for nuclei becomes large and it is difficult to constrain the value of the correction.

\subsection{Systematic uncertainties}
\label{sec:syst}

A summary of the systematic uncertainties for all the measurements is reported in Tab.~\ref{tab:systematics}. Values are provided for low (1.5~GeV$/c$) and high (4~GeV$/c$) transverse momentum. Where the systematic uncertainty differs between matter and antimatter, the latter is reported within brackets. The first source of systematic uncertainty is related to track selection. This source takes into account the imprecision in the description of the detector response in the MC simulation. The uncertainties are evaluated by varying the relevant selection criteria, as done in Ref.~\cite{deuteron_pp_7TeV}. It is worth mentioning that at low \pt uncertainties are generally larger for matter than for antimatter, due to the increasing number of secondary nuclei selected when loosing the selection on the DCA. It is one of the main sources. The second source is related to signal extraction. It is evaluated by changing the fit function used to evaluate the raw yield or, when the direct count is used, by varying the interval in which the count is performed (see Ref.~\cite{deuteron_pp_7TeV} and Ref.~\cite{deuteron_pp_13TeV} for further details). Its value increases with transverse momentum. The effect of the incomplete knowledge of the material budget of the detector is evaluated by comparing different MC simulations in which the material budget is varied by $\pm$4.5\%. This value corresponds to the uncertainty on the determination of the material budget by measuring photon conversions~\cite{ALICEperformance}. Similarly, the limited precision in the measurements of inelastic cross sections of (anti)nuclei with matter is a source of systematic uncertainty. It is evaluated by comparing experimental measurements of the inelastic cross section with the values implemented in GEANT4, following the same approach used in Ref.~\cite{antideuteronInelCS}.
For antihelion, the difference between the momentum-dependent inelastic cross sections implemented in GEANT3~\cite{GEANT3} and GEANT4 is also considered. This contribution is maximum (3$\%$) for $p_{\rm T} < 1.5$ GeV/$c$ and decreases to a negligible level going to higher $p_{\rm T}$.
Finally, the last sources of systematic uncertainty are related to the matching of the tracks between ITS and TPC and between TPC and TOF. They are evaluated from the difference between the ITS--TPC (TPC--TOF) matching in data and MC.

\begin{table}[ht]
\centering
\caption{Summary of the main contributions to the systematic uncertainties for all the particle species under study at $p_{\mathrm{T}} = 1.5$~GeV/$c$ and at $p_{\mathrm{T}} = 4$~GeV/$c$. Values in brackets refer to antiparticles. If they are not present, the systematic uncertainty is the same for particles and antiparticles. A dash symbol is used where the uncertainty from the corresponding source is not applicable. More details about the sources of the uncertainties can be found in the text.}
\label{tab:systematics}
\vspace{10pt}
\renewcommand{\arraystretch}{1.5}
\resizebox{\textwidth}{!}{
\begin{tabular}{|c|cccc|ccc|}
\hline
\multirow{2}{*}{Source} & \multicolumn{4}{c|}{$p_{\mathrm{T}} = 1.5$~GeV$/c$}                                                                                   & \multicolumn{3}{c|}{$p_{\mathrm{T}} = 4$~GeV$/c$}                                               \\
                        & p ($\overline{\mathrm{p}}$) & d ($\overline{\mathrm{d}}$) & $^3$He ($^3\overline{\mathrm{He}}$) & $^3$H ($^3\overline{\mathrm{H}}$) & p ($\overline{\mathrm{p}}$) & d ($\overline{\mathrm{d}}$) & $^3$He ($^3\overline{\mathrm{He}}$) \\ \hline
Track selection         & 3\%                         & 1\%                  & 14\% (10\%)                         & 14\% (10\%)                       & 3\%                         & 2\%                         & 10\% (7\%)                          \\
Signal extraction       & \textless1\%                & 3\%                         & 13\% (\textless{}1\%)               & \textless 1\%                     & 5\%                         & 2\%                         & \textless 1\%                       \\
Material budget         & 1\%                         & 1\%                         & 2\%                                 & 2\%                               & 1\%                         & \textless 1\%               & 2\%                                 \\
Hadronic interaction    & 1\% (2\%)                   & 2\% (6\%)                        & 1\% (2\%)                           & 9\% (6\%)                         & 1\% (2\%)                   & 2\% (6\%)                   & 1\% (1\%)                           \\
ITS--TPC matching        & 2\%                         & 2\%                         & 2\%                                 & 2\%                               & 2\%                         & 2\%                         & 2\%                                 \\
TPC--TOF matching        & 3\%                         & 2\%                         & $-$                                   & 3\%                               & 3\%                         & 2\%                         & $-$                                   \\ \hline
Total                   & 6\% (7\%)                   & 4\% (7\%)                   & 22\% (11\%)                         & 20\% (17\%)                       & 7\% (8\%)                   & 4\% (7\%)                   & 8\%                                \\ \hline
\end{tabular}
}
\end{table}

\FloatBarrier
\section{Results and discussion}
\label{sec:results}

\begin{figure}[ht]
\centering
    \includegraphics[width=\textwidth]{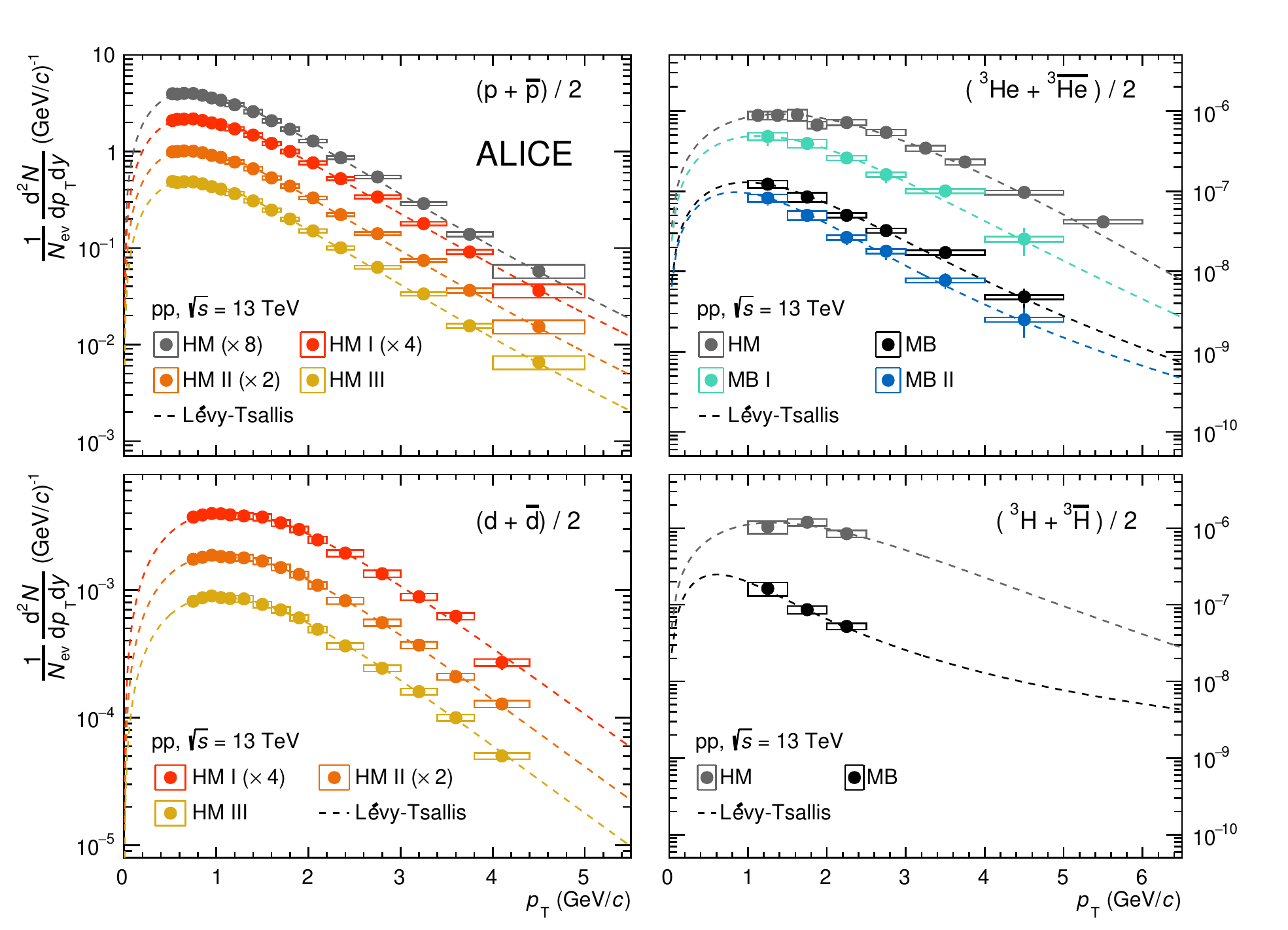}
    \caption{Transverse-momentum spectra of (anti)protons, (anti)deuterons, (anti)helion, and (anti)triton, measured in HM and MB pp collisions at $\sqrt{s}~=~$13~TeV at midrapidity ($|y| < 0.5)$. The results are shown in the multiplicity classes reported in Tab.~\ref{tab:multiplicity}. Vertical bars and boxes represent statistical and systematic uncertainties, respectively. The dashed lines are individual fits with a L\'evy-Tsallis function.}
    \label{fig:spectra}
\end{figure}

The production spectra for all the species under study are shown in Fig.~\ref{fig:spectra}. The multiplicity classes used for this measurement are reported in Tab~\ref{tab:multiplicity}, together with the corresponding \pt-integrated yields. (Anti)protons, (anti)deuterons and (anti)helions are fitted with a L\'evy-Tsallis function~\cite{Tsallis}, which is used to extrapolate the yields in the unmeasured \pt region. For (anti)triton, the fit parameters (except for the mass and the normalisation) are fixed to those of (anti)helion, due to the few data points available. The extrapolation amounts to about 20\% of the total \pt-integrated yield for (anti)protons and (anti)deuterons, about 30\% for (anti)helions, and about 50\% for (anti)tritons. Alternative fit functions such as a simple exponential depending on \mt, a Boltzmann function or a Blast-wave function~\cite{BlastWave1,BlastWave2,BlastWave3}, are used to evaluate the systematic uncertainty on the \pt-integrated yield as done in Ref.~\cite{deuteron_pp_13TeV,deuteron_pPbALICE,3He_pPb}. This uncertainty varies between 0.5\% and 3\% for protons, deuterons and helions in the HM analysis. For tritons it is around 8\% due to the narrower \pt coverage.
In the MB analysis, it is around 8\% (14\%) for helions (tritons).

\begin{table}
\centering
\caption{Multiplicity classes for the different measurements, with the corresponding multiplicity $\langle$d$N_{\mathrm{ch}}/$d$\eta\rangle_{|\eta_{\mathrm{lab}}| < 0.5}$, and \pt-integrated yields d$N/$d$y$ for the different species (average for particle and antiparticle). For $\langle$d$N_{\mathrm{ch}}/$d$\eta\rangle_{|\eta_{\mathrm{lab}}| < 0.5}$, only the systematic uncertainty is reported, because the statistical one is negligible. For d$N/$d$y$, the first uncertainty is statistical and the second is systematic.}
\label{tab:multiplicity}
\vspace{10pt}
\renewcommand{\arraystretch}{1.5}
\resizebox{\textwidth}{!}{
\begin{tabular}{|c|c|cccc|}
\hline
\multirow{2}{*}{Multiplicity} & \multirow{2}{*}{$\langle$d$N_{\mathrm{ch}}/$d$\eta\rangle_{|\eta_{\mathrm{lab}}| < 0.5}$} & \multicolumn{4}{c|}{d$N/$d$y$}                                                                                                                                \\
                                    &                                                    & p                          & d                                         & ${}^{3}$He                               & ${}^{3}$H                                \\ \hline
HM                                  & 31.5 $\pm$ 0.3                                     & 0.80 $\pm$ 0.01 $\pm$ 0.05 &                                           & (23.3 $\pm$ 1 $\pm$ 3) $\times 10^{-7}$    & (25 $\pm$ 2 $\pm$ 4) $\times 10^{-7}$      \\ \hline
HM I                                & 35.8 $\pm$ 0.5                                     & 0.91 $\pm$ 0.01 $\pm$ 0.05 & (22.1 $\pm$ 0.1 $\pm$ 1.4) $\times 10^{-4}$ &                                          &                                          \\
HM II                               & 32.2 $\pm$ 0.4                                     & 0.83 $\pm$ 0.01 $\pm$ 0.05 & (19.8 $\pm$ 0.1 $\pm$ 1.3) $\times 10^{-4}$ &                                          &                                          \\
HM III                              & 30.1 $\pm$ 0.4                                     & 0.77 $\pm$ 0.01 $\pm$ 0.04 & (18.4 $\pm$ 0.1 $\pm$ 1.1) $\times 10^{-4}$ &                                          &                                          \\ \hline
MB                                  & 6.9  $\pm$ 0.1                                     &                            &                                           & (2.4 $\pm$ 0.3 $\pm$ 0.4) $\times 10^{-7}$ & (1.7 $\pm$ 0.3 $\pm$ 0.4) $\times 10^{-7}$ \\ \hline
MB I                                & 18.7 $\pm$ 0.3                                     &                            &                                           & (11 $\pm$ 2 $\pm$ 2) $\times 10^{-7}$      &                                          \\
MB II                               & 6.0  $\pm$ 0.2                                     &                            &                                           & (1.5 $\pm$ 0.2 $\pm$ 0.3) $\times 10^{-7}$ &                                          \\ \hline
\end{tabular}
}
\end{table}

\subsection{Coalescence parameter as a function of transverse momentum}

In the coalescence model, the production probability of a nucleus with mass number $A$ is proportional to the coalescence parameter $B_A$, defined as

\begin{equation}
    B_{A}\left(p_{\mathrm{T}}^{\mathrm{p}}\right) = \frac{1}{2\pi p_{\mathrm{T}}^{\mathrm{A}}}\frac{\mathrm{d}^2N_{\mathrm{A}}}{\mathrm{d}y\mathrm{d}p^{\mathrm{A}}_{\mathrm{T}}} \; \bigg/ \left(\frac{1}{2\pi p_{\mathrm{T}}^{\mathrm{p}}}\frac{\mathrm{d}^2N_{\mathrm{p}}}{\mathrm{d}y\mathrm{d}\pt^{\mathrm{p}}}\right)^{A} ,
\end{equation}

where the labels $\mathrm{p}$ and $\mathrm{A}$ refer to protons and the (anti)nucleus with mass-number $A$, respectively. The invariant spectra of the $\text{(anti)protons}$ are evaluated at the transverse momentum of the (anti)nucleus, divided by the mass-number $A$. For (anti)deuterons, for example, $B_{A} = B_{2}$ and $p_{\mathrm{T}}^{\mathrm{p}} = p_{\mathrm{T}}^{\mathrm{A}} / A = p_{\mathrm{T}}^{\mathrm{d}} / 2$.

\begin{figure}[ht]
    \centering
    \captionsetup[subfigure]{justification=centering}
    \begin{subfigure}[t]{0.49\textwidth}
        \centering
        \includegraphics[width=\textwidth]{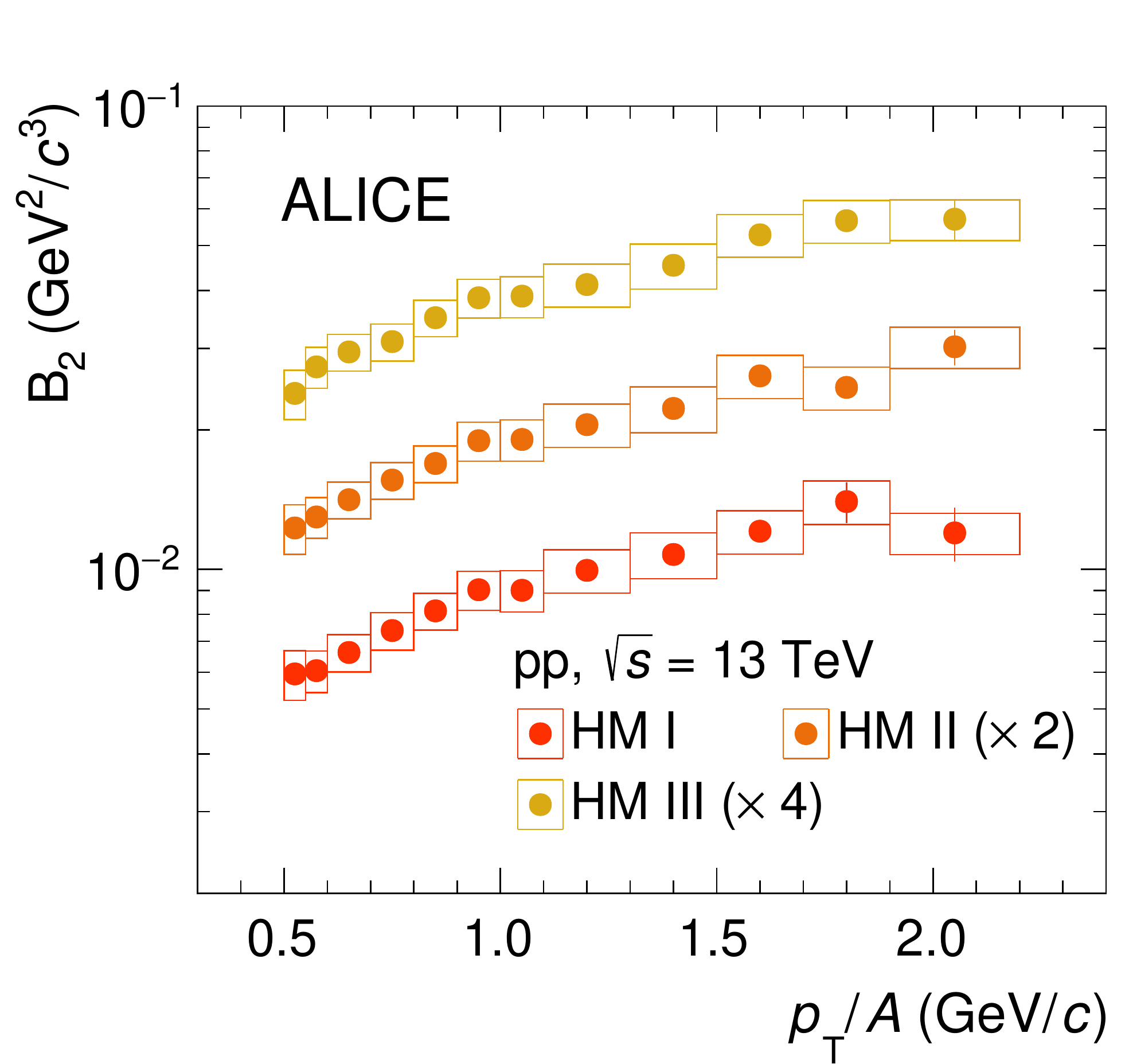}
        \caption{(Anti)deuterons}
        \label{fig:B2data}
    \end{subfigure}
    \begin{subfigure}[t]{0.49\textwidth}
        \centering
        \includegraphics[width=\textwidth]{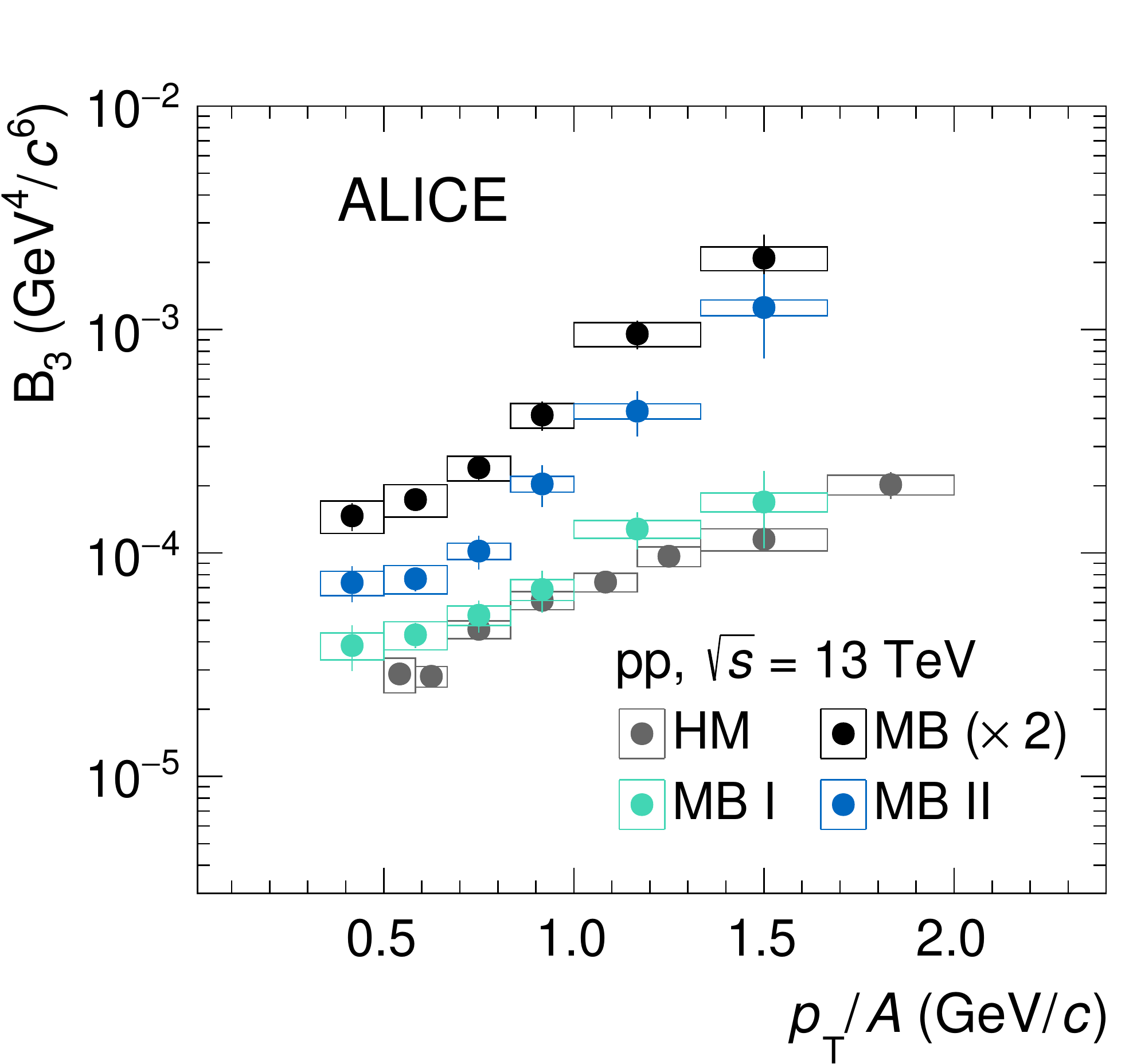}
        \caption{(Anti)helions}
        \label{fig:B3data}
    \end{subfigure}
    \caption{Coalescence parameters $B_{2}$ (a) and $B_{3} $ (b) as a function of $p_{\mathrm{T}}/A$ for the multiplicity classes reported in Tab.~\ref{tab:multiplicity}. Vertical bars and boxes represent statistical and systematic uncertainties, respectively.}
    \label{fig:BA_data}
\end{figure}

\begin{figure}[ht]
    \centering
    \captionsetup[subfigure]{justification=centering}
    \begin{subfigure}[t]{0.49\textwidth}
        \centering
        \includegraphics[width=\textwidth]{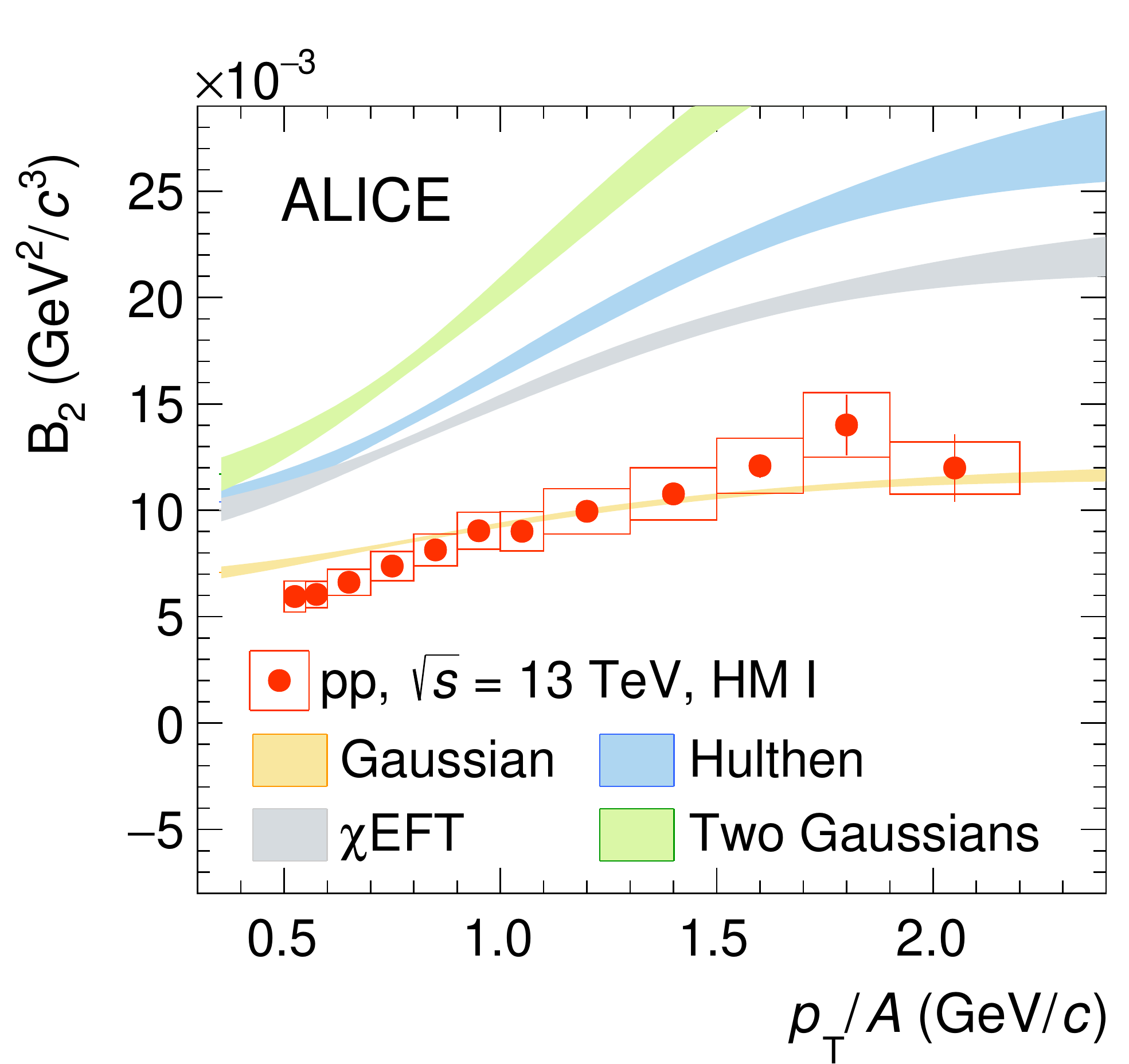}
        \caption{(Anti)deuterons}
        \label{fig:B2theory}
    \end{subfigure}
    \begin{subfigure}[t]{0.49\textwidth}
        \centering
        \includegraphics[width=\textwidth]{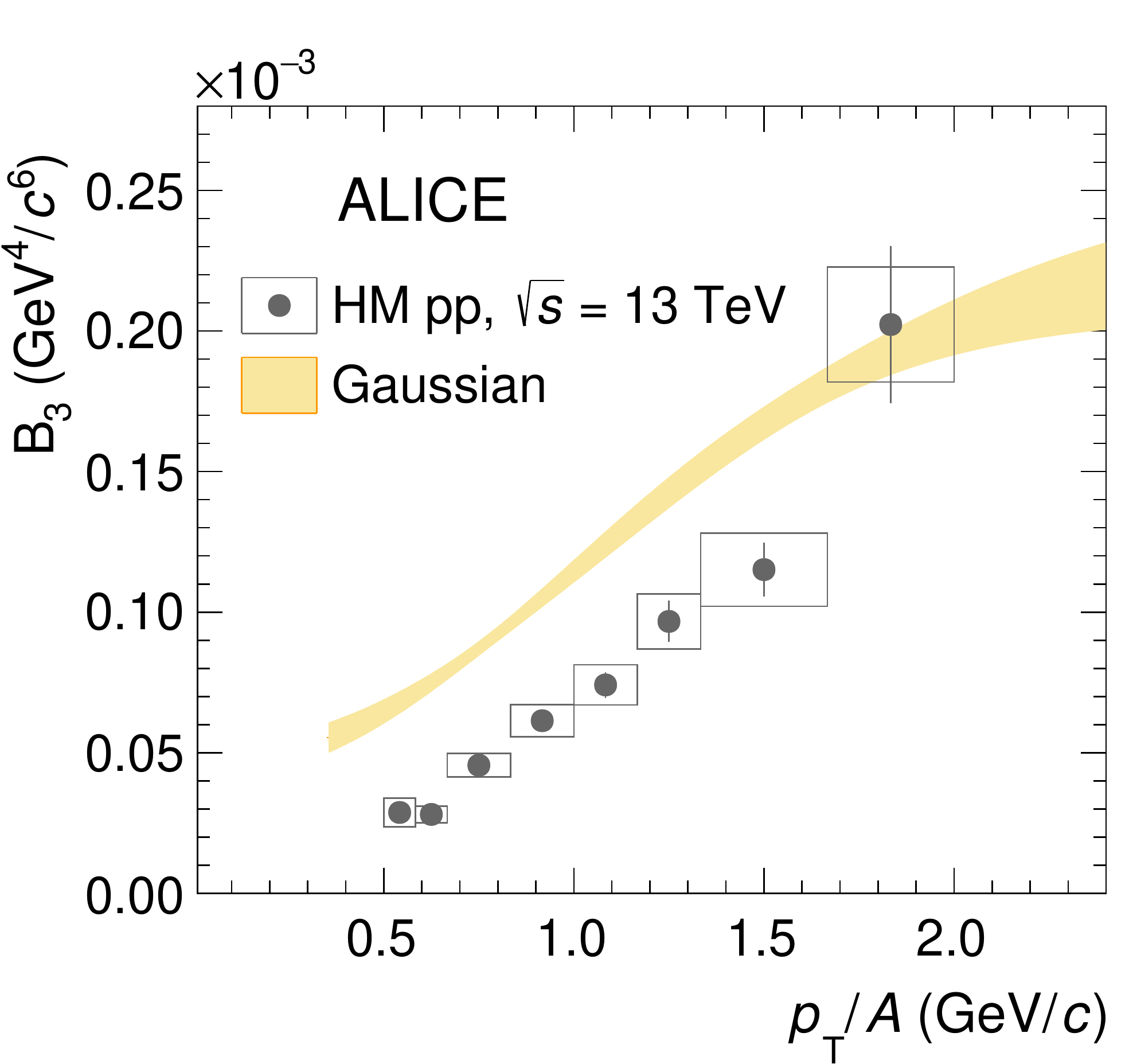}
        \caption{(Anti)helions}
        \label{fig:B3theory}
    \end{subfigure}
    \caption{Comparison between measurements and theoretical predictions for the coalescence parameters $B_{2}$ for (anti)deuterons (a) and $B_{3} $ for (anti)helions (b) as a function of $p_{\mathrm{T}}/A$. Vertical bars and boxes represent statistical and systematic uncertainties, respectively. Theoretical predictions are obtained using different wave functions to describe nuclei: Gaussian (yellow), Hulthen (blue), $\chi \text{EFT}$ (gray) and two Gaussians (green).}
    \label{fig:BA_theory}
\end{figure}

The coalescence parameters for (anti)deuterons ($B_2$) and for (anti)helions ($B_3$) are shown in Fig.~\ref{fig:BA_data} as a function of $\pt/A$ for different multiplicity classes. Some of the measurements are scaled for better visibility (see the legend), but for $B_2$ the three curves are consistent with each other. A clear increase of both $B_2$ and $B_3$ with increasing $\pt/A$ is observed. Previous measurements of $B_2$ in pp~\cite{deuteron_pp_13TeV, deuteron_pp_7TeV} and p--Pb~\cite{deuteron_pPbALICE} collisions indicated an almost flat trend with $\pt/A$. However, in Ref.~\cite{deuteron_pp_13TeV} and in Ref.~\cite{deuteron_pp_7TeV} it was shown that even though $B_2$ evaluated in multiplicity classes was flat, $B_2$ evaluated in the multiplicity-integrated sample showed a rise with $\pt/A$. The trend shown in Ref.~\cite{deuteron_pp_7TeV} is a consequence of the mathematical definition of $B_2$ and of the hardening of the proton spectra. Given the narrow multiplicity intervals used in the present measurement, the significant rise of the coalescence parameters with $\pt/A$ cannot be attributed to effects originating from a different hardening of the (anti)proton and (anti)nuclei spectra within these multiplicity intervals~\cite{deuteron_pp_7TeV}.

The measurement of the coalescence parameter as a function of transverse momentum is compared with predictions from the coalescence model, using different nuclei wave functions~\cite{coalescenceFemto} and the precise measurement of the source radii for the same data set~\cite{sourceSizeHMpp}. In the case of (anti)deuterons, the following wave functions are used: single and double Gaussian~\cite{alternativeCoalescence}, Hulthen~\cite{Coalescence3}, and a function obtained from chiral Effective Field Theory ($\chi \text{EFT}$) of order N4LO with a cutoff at 500~MeV~\cite{chiEFT}. For (anti)helion, only calculations using a  Gaussian wave function are currently available, because the general recipe used for $B_2$ cannot be extended to $B_3$ but new calculations \textit{ab initio} are needed. These wave functions and more details about the adopted theoretical models can be found in Appendix~\ref{app:theory}.
In the coalescence model, the coalescence parameter depends on the radial extension of the particle emitting source~\cite{coalescenceFemto}. The source radius is measured in HM pp collisions at $\sqrt{s} = $~13~TeV by ALICE using p--p and p--$\Lambda$ correlations as a function of the mean transverse mass $\langle m_{\rm T}\rangle$ of the pair ~\cite{sourceSizeHMpp}. In Ref.~\cite{sourceSizeHMpp} two different measurements of the source radius are reported, $r_{\mathrm{effective}}$ and $r_{\mathrm{core}}$, respectively. The difference between the two is that $r_{\mathrm{core}}$ takes into account the contributions coming from the strong decay of resonances by subtracting them. It is shown that $r_{\mathrm{core}}$ is universal, since it could describe simultaneously p--p and p--$\Lambda$ correlations. In this analysis, $r_{\mathrm{core}}$ is used. The difference between $r_{\mathrm{core}}$ and $r_{\mathrm{effective}}$ is small: $B_2$ is on average 7\% smaller using $r_{\mathrm{effective}}$, while $B_3$ is on average 20\% smaller, due to the stronger dependence on the system size for $A>$2 (see Eq.~\ref{eq:BA_gauss}). For $B_2$, only the HM I class is considered, because the $B_2$ values in the three multiplicity classes are compatible.

The data in Ref.~\cite{sourceSizeHMpp} are parameterised as $r_{\rm source} (\langle m_{\rm T}\rangle) = c_0 + \mathrm{exp}(c_1+c_2\langle m_{\rm T}\rangle)$, with $c_i$ free parameters, to map the transverse-mass to the source radius. The value of \pt corresponding to $m_{\mathrm{T}}$ is taken from $m_{\rm T}=\sqrt{m^{2}_{\rm p} + (p_{\mathrm{T}}/A)^{2}}$, where $m_{\rm p}$ is the proton mass. The radius of the deuteron and $^{3}$He are taken from Ref.~\cite{coalescenceBelliniKalweit}.
The coalescence predictions are shown in comparison with the data in Fig.~\ref{fig:BA_theory}. Bands represent the uncertainty propagated from the measurement of the source radius. In the case of $B_2$, the Gaussian wave function provides the best description of the data, even though the Hulthen wave function is favoured by low-energy scattering experiments~\cite{Phillips_1959}. The other wave functions are significantly larger than the measurement. For the $B_3$, the coalescence predictions using a Gaussian wave function of helion are above the data by almost a factor of 2 except for the last \pt interval, which is consistent with the measured $B_3$ within the uncertainties. In the future, a systematic investigation of the coalescence parameter $B_3$ using different wave functions, in the context of the coalescence model, will gauge the potential of coalescence measurements to further constrain the wave function of helion. This technique can be used in a more general context to obtain information on the internal wave function of more complex (hyper)nuclei, such as alpha ($^{4}$He) and hypertriton.

\begin{figure}
\centering
\includegraphics[width=0.8\textwidth]{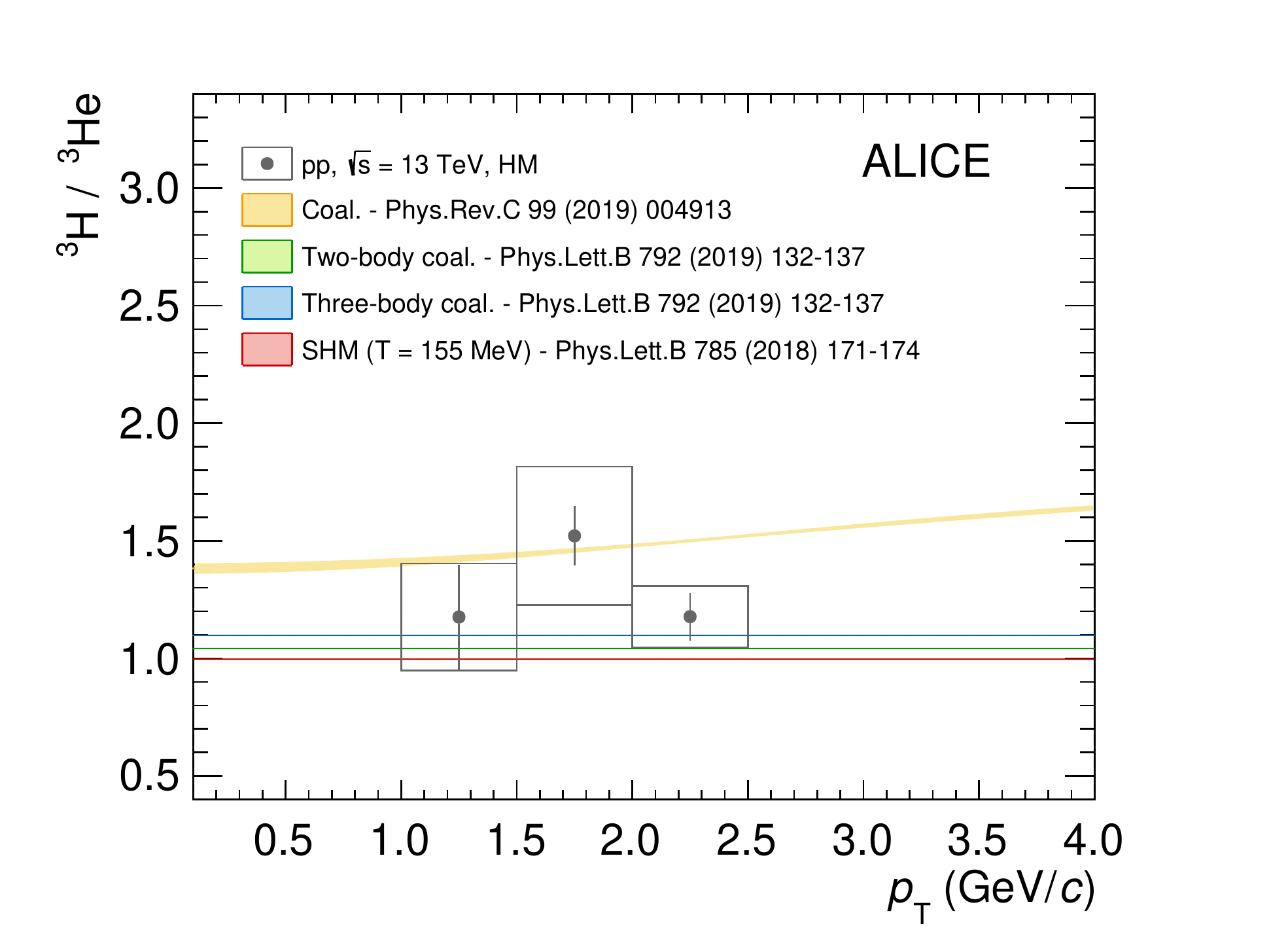}
\caption{Ratio between the \pt spectra of triton and helion for the HM data sample. Vertical bars and boxes represent statistical and systematic uncertainties, respectively. The measurements are compared with the prediction of thermal (red) and coalescence models from Ref.~\cite{coalescence_correlations} (yellow) and Ref.~\cite{coalescenceSmallSystems} (green and blue).}
\label{fig:tritonToHelium}
\end{figure}

\subsection{Ratio between triton and helion yields}

The statistical hadronisation and coalescence models predict different yields of nuclei with similar masses but different radii. To test the production model, the ratio of triton and helion is measured as a function of \pt for HM pp collisions (Fig.~\ref{fig:tritonToHelium}) and compared with the model predictions. Two different versions of coalescence are considered, based on Ref.~\cite{coalescenceSmallSystems} and Ref.~\cite{coalescence_correlations}, respectively. The main difference between the two models concerns the source size $R$: while in Ref.~\cite{coalescenceSmallSystems} the value of $R$ is constrained from the parameters of a thermal fit, in Ref.~\cite{coalescence_correlations} $R$ is an independent variable, for which the aforementioned \pt dependence has been taken into account. In the former approach, $R$ is about a factor of 2 larger than in the latter, determining very different predictions. The coalescence model predicts a slightly larger yield of triton as compared to helion due to its smaller nuclear radius. In the statistical hadronisation model, the yield ratio between these two nuclei is given by $\exp(-\Delta m/T_{\rm chem})$, where $\Delta m$ is the mass difference between triton and helion, taken from Ref.~\cite{massDifference}, and $T_{\rm chem}$ the chemical freeze-out temperature. For the latter, $T_{\rm chem}=155$~MeV is used, as done in the canonical statistical model~\cite{CSM}. Given the small mass difference, the statistical hadronisation model predicts a ratio which is very close to unity. The precision of the present data prevents distinguishing between the models. The $^3$H/$^3$He ratio will be measured with higher precision in Run 3~\cite{pnRun3} of the LHC. Indeed, the new ITS, which is characterised by a low material budget, will reduce the systematic uncertainty related to track reconstruction~\cite{ITSU}. Moreover, with a better description of the nuclear absorption cross section, it will be possible to reduce the corresponding systematic uncertainties.

\begin{figure}[ht]
    \centering
    \captionsetup[subfigure]{justification=centering}
    \begin{subfigure}[t]{0.49\textwidth}
        \centering
        \includegraphics[width=\textwidth]{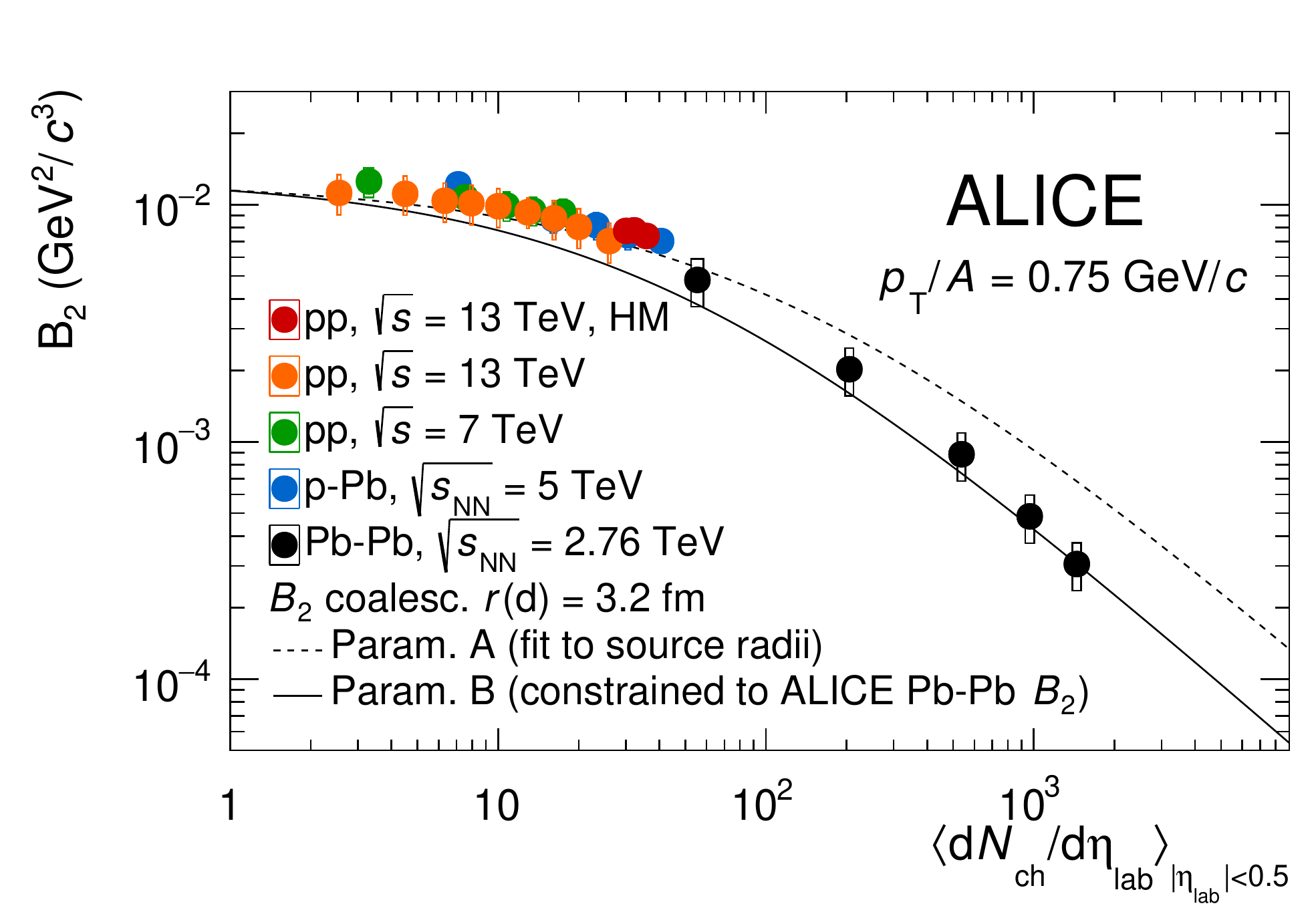}
        \caption{(Anti)deuterons}
        \label{fig:B2mult}
    \end{subfigure}
    \begin{subfigure}[t]{0.49\textwidth}
        \centering
        \includegraphics[width=\textwidth]{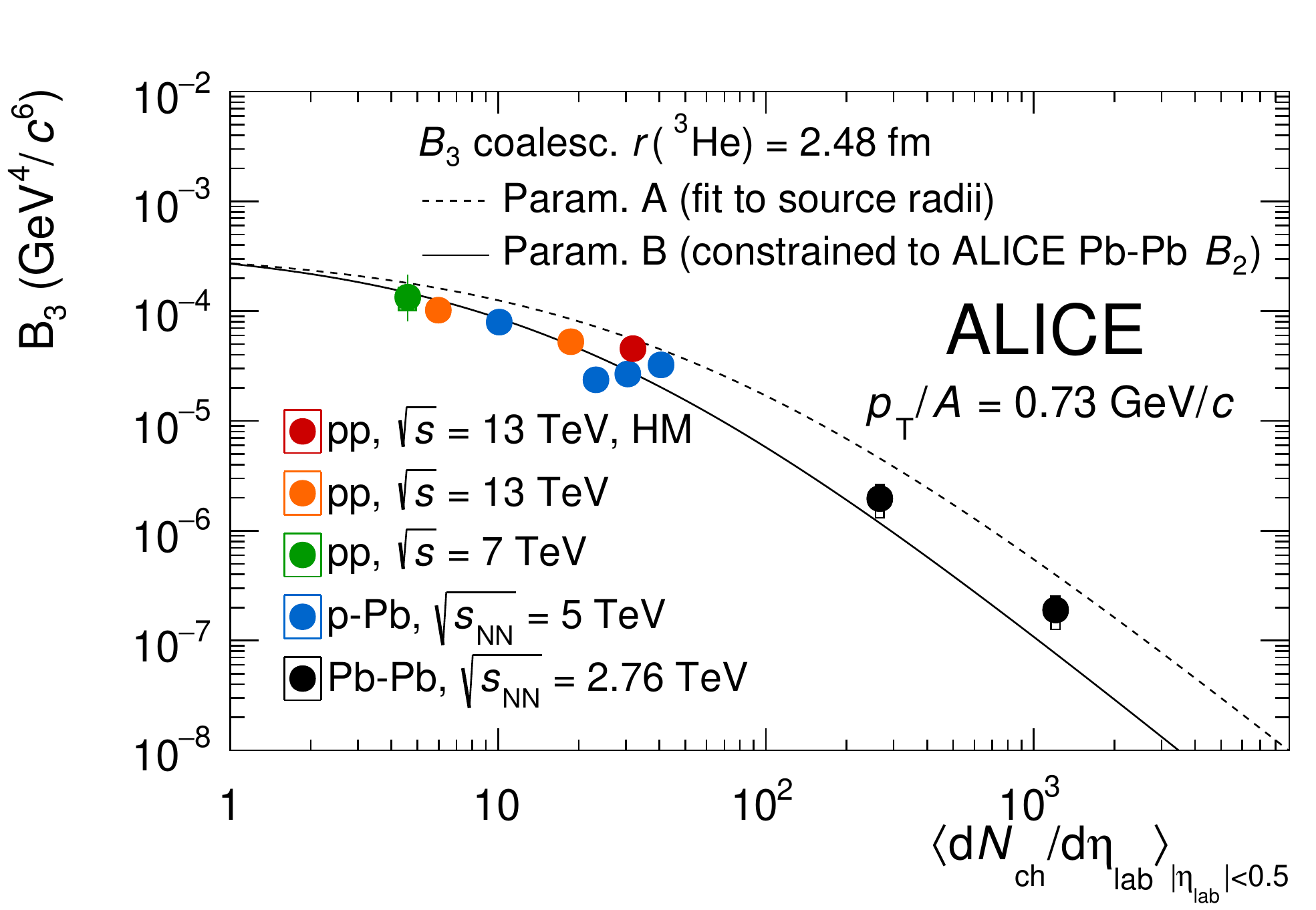}
        \caption{(Anti)helions}
        \label{fig:B3mult}
    \end{subfigure}
    \caption{(a): $B_2$ at $p_{\mathrm{T}}/A = $ 0.75~GeV/\textit{c} as a function of multiplicity in HM pp collisions at $\sqrt{s}~=~13$~TeV, in MB pp collisions at $\sqrt{s}~=~13$~TeV~\cite{deuteron_pp_13TeV} and at $\sqrt{s}~=~7$~TeV~\cite{deuteron_pp_7TeV}, in p--Pb collisions at $\snn~=~5.02$~TeV~\cite{deuteron_pPbALICE}, and in Pb--Pb collisions at $\snn~=~2.76$~TeV~\cite{nuclei_pp_PbPb}. (b): $B_3$ at $p_{\mathrm{T}}/A = $ 0.73~GeV$/$\textit{c} as a function of multiplicity in HM pp collisions at $\sqrt{s}~=~13$~TeV, in MB pp collisions at $\sqrt{s}~=~13$~TeV, in p--Pb collisions at $\snn~=~5.02$~TeV~\cite{deuteron_pPbALICE}, and in Pb--Pb collisions at $\snn~=~2.76$~TeV~\cite{nuclei_pp_PbPb}. Vertical bars and boxes represent statistical and systematic uncertainties, respectively. The two lines are theoretical predictions based on two different parameterisations of the source radius.}
    \label{fig:BA_mult}
\end{figure}

\subsection{Coalescence parameter as a function of multiplicity}

The evolution of $B_A$ with multiplicity \avdndeta provides an insight on the dependence of the production mechanisms of light (anti)nuclei. Fig.~\ref{fig:BA_mult}(a) shows $B_2$ as a function of \avdndeta for different collision systems and energies at $\pt~=~0.75$~GeV$/c$ and $B_3$ at $\pt~=~0.73$~GeV$/c$. The measurements are compared with the theoretical predictions from Ref.~\cite{coalescenceBelliniKalweit}, using $r(\mathrm{d}) = 3.2$~fm and $r(^3\mathrm{He}) = 2.48$~fm as deuteron and helion radii, respectively. Two different parameterisations (named A and B in the following) of the system radius as a function of multiplicity are used. Parameterisation A is based on a fit to the ALICE measurements of system radii $R$ from femtoscopic measurement as a function of multiplicity~\cite{oldHBTradii}. In parameterisation B, the relation between the system radius and the multiplicity is fixed to reproduce the $B_2$ of deuterons in \PbPb collisions at $\snn=2.76$~TeV in the centrality class 0--10\% (see Ref.~\cite{coalescenceBelliniKalweit} for more details). The $B_2$ measurement in HM pp collisions agrees with observations in p--Pb collisions at $\sqrt{s_{\mathrm{NN}}} =$~5~TeV~\cite{deuteron_pPbALICE}  at similar multiplicity and confirms the trend observed in all the previous measurements. This measurement further strengthen the idea of a production mechanism that depends only on the multiplicity and not on the collision system nor the centre-of-mass energy. Similarly, Fig.~\ref{fig:BA_mult}(b) shows the evolution of $B_3$ as a function of multiplicity. The measurements are compared with the theoretical prediction from Ref.~\cite{coalescenceBelliniKalweit}, using the same two paremeterisations as for $B_2$. Also in this case, the coalescence model qualitatively describes the trend but fails in accurately describing the measurements in the whole multiplicity range. For both $B_2$ and $B_3$, one reason could be that multiplicity is not a perfect proxy for the system size, because for each multiplicity the source radius depends also on the transverse momentum of the particle of interest, as shown in Fig.~\ref{fig:BA_data}. In the future, it is important to have more measurements of the source radius as a function of \mt for the different multiplicity classes in order to test the agreement between the models and the $B_A$ measurement over the whole multiplicity range.

\subsection{Ratio between integrated yields of nuclei and protons as a function of multiplicity}
\begin{figure}[ht]
    \centering
    \captionsetup[subfigure]{justification=centering}
    \begin{subfigure}[t]{0.49\textwidth}
        \centering
        \includegraphics[width=\textwidth]{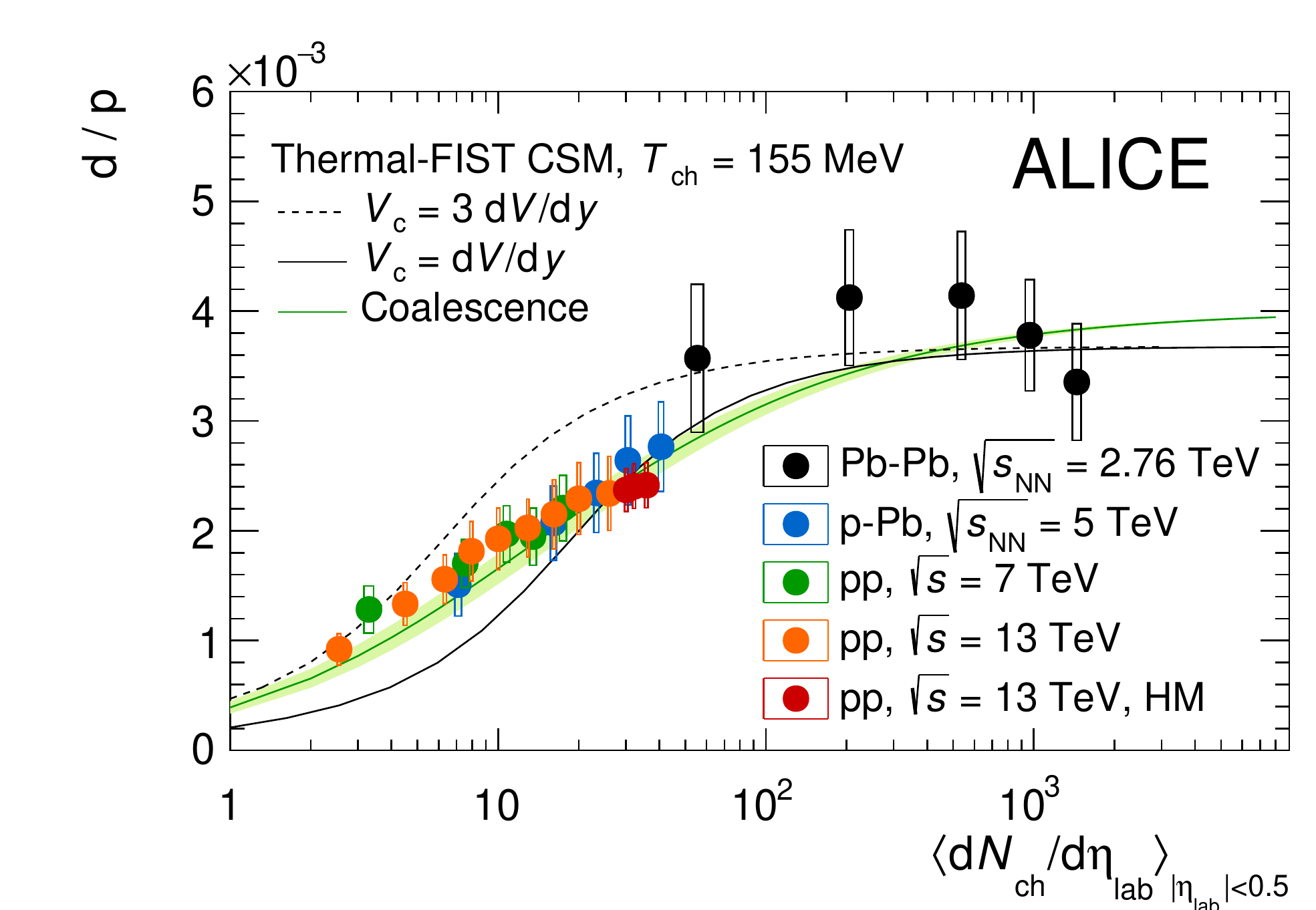}
        \caption{(Anti)deuterons}
        \label{fig:DoP}
    \end{subfigure}
    \begin{subfigure}[t]{0.49\textwidth}
        \centering
        \includegraphics[width=\textwidth]{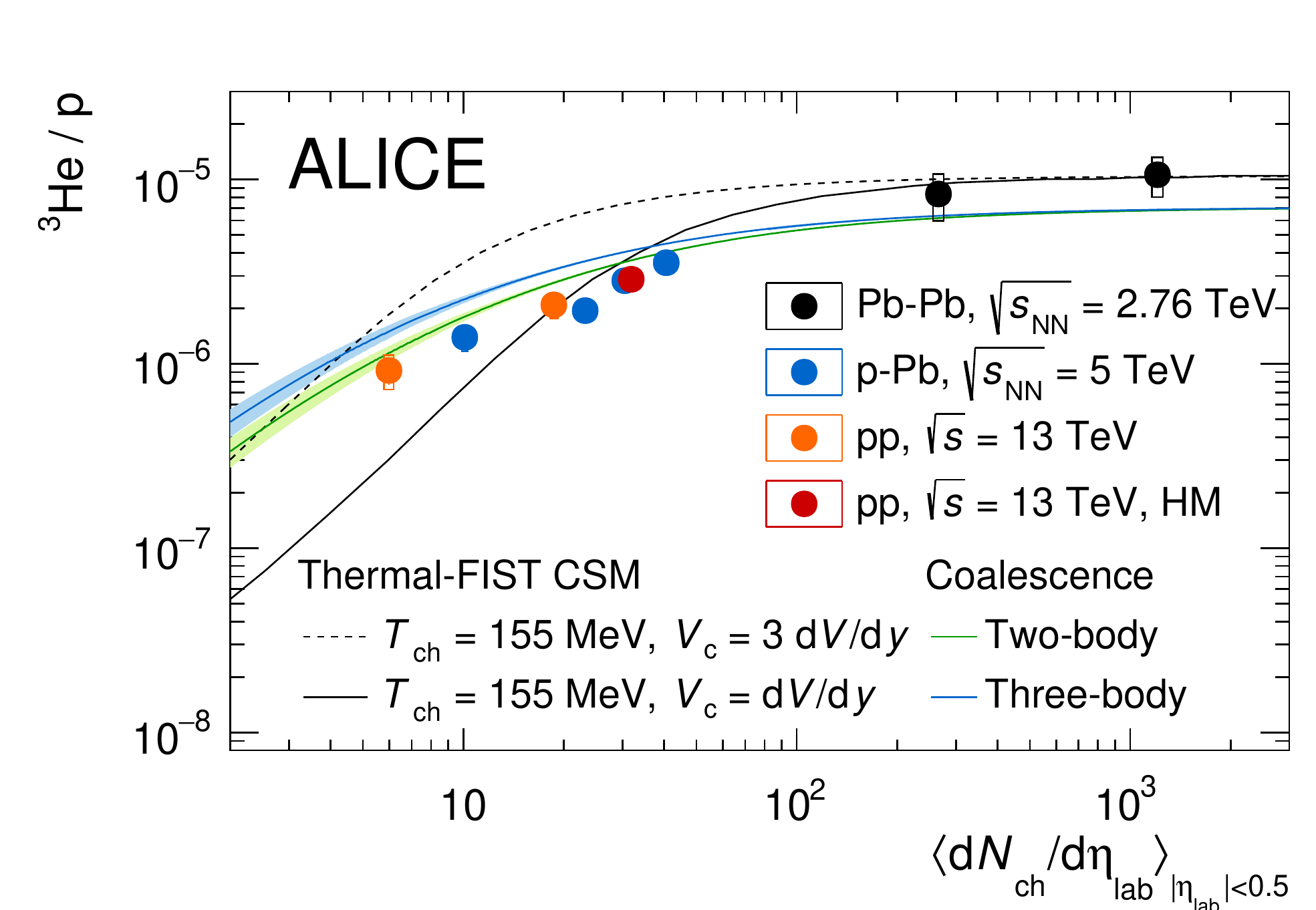}
        \caption{(Anti)helions}
        \label{fig:HEoP}
    \end{subfigure}
    \caption{Ratio between the \pt-integrated yields of nuclei and protons as a function of multiplicity for (anti)deuterons (a) and (anti)helions (b). Measurements are performed in HM pp collisions at $\sqrt{s}~=~13$~TeV, in MB pp collisions at $\sqrt{s}~=~13$~TeV~\cite{deuteron_pp_13TeV} and at $\sqrt{s}~=~7$~TeV~\cite{deuteron_pp_7TeV}, in p--Pb collisions at $\snn~=~5.02$~TeV~\cite{deuteron_pPbALICE,3He_pPb}, and in Pb--Pb collisions at $\snn~=~2.76$~TeV~\cite{nuclei_pp_PbPb}. Vertical bars and boxes represent statistical and systematic uncertainties, respectively. The two black lines are the theoretical predictions of the Thermal-FIST CSM~\cite{CSM} for two sizes of the correlation volume $V_{\mathrm{C}}$. For (anti)deuterons, the green line represents the expectation from a coalescence model~\cite{coalescenceSmallSystems}. For (anti)helion, the blue and green lines represent the expectations from a two-body and three-body coalescence model, respectively~\cite{coalescenceSmallSystems}.}
    \label{fig:AoP_mult}
\end{figure}

The measurements of the ratios between the \pt-integrated yields of nuclei and protons as a function of multiplicity are shown.
Figure~\ref{fig:AoP_mult} shows the measurement for different collision systems and energies for deuterons (d$/$p) and helions (${}^{3}$He$/$p), on the left and right panels, respectively. The new measurements complement the existing picture and are consistent with the global trend obtained from previous measurements ~\cite{nuclei_pp_PbPb,deuteron_pp_7TeV,deuteron_PbPb_276TeV,nuclei_pp,4He_PbPb,deuteron_pPbALICE,3He_pPb,deuteron_pp_13TeV}, for both d$/$p and {${}^{3}$He$/$p}: the ratio increases as a function of multiplicity and eventually saturates at high multiplicities. This trend can be interpreted as a consequence of the interplay between the evolution of the yields and of the system size with multiplicity. The measurements are compared with the prediction of both Thermal-FIST Canonical Statistical Model (CSM)~\cite{CSM} and coalescence model~\cite{coalescenceSmallSystems}. The predictions from CSM suggest correlation volumes $V_{\mathrm{C}}$ between 1 and 3 units of rapidity. However, the measurement of the proton-to-pion (p/$\pi$) ratio is better described by a correlation volume of 6 rapidity units~\cite{vanillaCSM}.
On the contrary, the coalescence model provides a better description of the data. For d$/$p, the agreement is good for the whole multiplicity range. For ${}^{3}$He$/$p, instead, there are more tensions between data and model in the multiplicity region corresponding to p--Pb and high-multiplicity pp collisions. Remarkably, the two-body coalescence appears to describe the data better than the three-body coalescence prediction.

%\FloatBarrier
\section{Summary}
\label{sec:Summary}

In this paper, the measurements of the production yields of (anti)nuclei in minimum bias and high-multiplicity pp collisions at $\sqrt{s} =$ 13~TeV are reported. 

A significant increase of the coalescence parameter $B_2$ with increasing $\pt/A$ is observed for the first time in pp collisions. Indeed, previous measurements in small collision systems were consistent with a flat trend within uncertainties. Given the very narrow multiplicity intervals used in the present measurement, this rising trend cannot be attributed to effects coming from a different hardening of the proton and deuteron spectra within the measured multiplicity intervals and thus points to some other physics effect. Moreover, the coalescence parameters are compared with theoretical calculations based on the coalescence approach using different internal wave functions of nuclei. This comparison was possible due to the availability of the measurement of the source radii in the same data sample. While the predictions for $B_2$ using a Gaussian approximation for the deuteron wave function are in very good agreement with the experimental results, for $B_3$ they overestimate the data by up to a factor of 2 at the lowest $\pt$. Updated theoretical calculations including also more complex wave functions for $^3$He would help in providing a better description of this measurement. 

The multiplicity evolution of the coalescence parameters for a fixed $\pt/A$ and of the ratios of integrated yields d$/$p and $^{3}$He$/$p are consistent with the global trend from previous measurements. Moreover, the d$/$p ratio is consistent with predictions from the coalescence model, while significant deviations are observed between the $^{3}$He$/$p and coalescence expectations at intermediate multiplicities. The canonical statistical model predictions provide a qualitative description of the particle ratios presented in this paper at low and intermediate multiplicities covered by pp and p--Pb collisions and are consistent with the data only in the grand-canonical limit (multiplicities covered by Pb--Pb collisions).

%\clearpage

%%%%%%%%%%%%%%%%%%%%%%%%%%%%%%%%
% end main text 
%%%%%%%%%%%%%%%%%%%%%%%%%%%%%%%%

%%%%% acknowledgements - handled by EB chairs 

%%%%% acknowledgements
\newenvironment{acknowledgement}{\relax}{\relax}
\begin{acknowledgement}
\section*{Acknowledgements}
% Version: 2021-08-25

The ALICE Collaboration would like to thank all its engineers and technicians for their invaluable contributions to the construction of the experiment and the CERN accelerator teams for the outstanding performance of the LHC complex.
The ALICE Collaboration gratefully acknowledges the resources and support provided by all Grid centres and the Worldwide LHC Computing Grid (WLCG) collaboration.
The ALICE Collaboration acknowledges the following funding agencies for their support in building and running the ALICE detector:
A. I. Alikhanyan National Science Laboratory (Yerevan Physics Institute) Foundation (ANSL), State Committee of Science and World Federation of Scientists (WFS), Armenia;
Austrian Academy of Sciences, Austrian Science Fund (FWF): [M 2467-N36] and Nationalstiftung f\"{u}r Forschung, Technologie und Entwicklung, Austria;
Ministry of Communications and High Technologies, National Nuclear Research Center, Azerbaijan;
Conselho Nacional de Desenvolvimento Cient\'{\i}fico e Tecnol\'{o}gico (CNPq), Financiadora de Estudos e Projetos (Finep), Funda\c{c}\~{a}o de Amparo \`{a} Pesquisa do Estado de S\~{a}o Paulo (FAPESP) and Universidade Federal do Rio Grande do Sul (UFRGS), Brazil;
Ministry of Education of China (MOEC) , Ministry of Science \& Technology of China (MSTC) and National Natural Science Foundation of China (NSFC), China;
Ministry of Science and Education and Croatian Science Foundation, Croatia;
Centro de Aplicaciones Tecnol\'{o}gicas y Desarrollo Nuclear (CEADEN), Cubaenerg\'{\i}a, Cuba;
Ministry of Education, Youth and Sports of the Czech Republic, Czech Republic;
The Danish Council for Independent Research | Natural Sciences, the VILLUM FONDEN and Danish National Research Foundation (DNRF), Denmark;
Helsinki Institute of Physics (HIP), Finland;
Commissariat \`{a} l'Energie Atomique (CEA) and Institut National de Physique Nucl\'{e}aire et de Physique des Particules (IN2P3) and Centre National de la Recherche Scientifique (CNRS), France;
Bundesministerium f\"{u}r Bildung und Forschung (BMBF) and GSI Helmholtzzentrum f\"{u}r Schwerionenforschung GmbH, Germany;
General Secretariat for Research and Technology, Ministry of Education, Research and Religions, Greece;
National Research, Development and Innovation Office, Hungary;
Department of Atomic Energy Government of India (DAE), Department of Science and Technology, Government of India (DST), University Grants Commission, Government of India (UGC) and Council of Scientific and Industrial Research (CSIR), India;
Indonesian Institute of Science, Indonesia;
Istituto Nazionale di Fisica Nucleare (INFN), Italy;
Japanese Ministry of Education, Culture, Sports, Science and Technology (MEXT), Japan Society for the Promotion of Science (JSPS) KAKENHI and Japanese Ministry of Education, Culture, Sports, Science and Technology (MEXT)of Applied Science (IIST), Japan;
Consejo Nacional de Ciencia (CONACYT) y Tecnolog\'{i}a, through Fondo de Cooperaci\'{o}n Internacional en Ciencia y Tecnolog\'{i}a (FONCICYT) and Direcci\'{o}n General de Asuntos del Personal Academico (DGAPA), Mexico;
Nederlandse Organisatie voor Wetenschappelijk Onderzoek (NWO), Netherlands;
The Research Council of Norway, Norway;
Commission on Science and Technology for Sustainable Development in the South (COMSATS) and Pakistan Atomic Energy Commission, Pakistan;
Pontificia Universidad Cat\'{o}lica del Per\'{u}, Peru;
Ministry of Education and Science, National Science Centre and WUT ID-UB, Poland;
Korea Institute of Science and Technology Information and National Research Foundation of Korea (NRF), Republic of Korea;
Ministry of Education and Scientific Research, Institute of Atomic Physics and Ministry of Research and Innovation and Institute of Atomic Physics, Romania;
Joint Institute for Nuclear Research (JINR), Ministry of Education and Science of the Russian Federation, National Research Centre Kurchatov Institute, Russian Science Foundation and Russian Foundation for Basic Research, Russia;
Ministry of Education, Science, Research and Sport of the Slovak Republic, Slovakia;
National Research Foundation of South Africa, South Africa;
Swedish Research Council (VR) and Knut \& Alice Wallenberg Foundation (KAW), Sweden;
European Organization for Nuclear Research, Switzerland;
Suranaree University of Technology (SUT), National Science and Technology Development Agency (NSDTA) and Office of the Higher Education Commission under NRU project of Thailand, Thailand;
Turkish Energy, Nuclear and Mineral Research Agency (TENMAK), Turkey;
National Academy of  Sciences of Ukraine, Ukraine;
Science and Technology Facilities Council (STFC), United Kingdom;
National Science Foundation of the United States of America (NSF) and United States Department of Energy, Office of Nuclear Physics (DOE NP), United States of America.    %%%%%%% done by webmaster team
In addition, individual groups and members have received support from European Research Council, European Union.
\end{acknowledgement}

%%%%%%%% Bibliography 
\bibliographystyle{utphys}    
\bibliography{bibliography}

\providecommand{\href}[2]{#2}\begingroup\raggedright\begin{thebibliography}{10}

\bibitem{hypertriton_PbPb_276}
{\bfseries ALICE} Collaboration, J.~Adam {\em et~al.},
  ``{$^{3}_{\Lambda}\mathrm H$ and $^{3}_{\bar{\Lambda}} \overline{\mathrm H}$
  production in Pb-Pb collisions at $\sqrt{s_{\rm NN}} =$ 2.76 TeV}'',
  \href{http://dx.doi.org/10.1016/j.physletb.2016.01.040}{{\em Phys. Lett.}
  {\bfseries B754} (2016) 360--372},
\href{http://arxiv.org/abs/1506.08453}{{\ttfamily arXiv:1506.08453 [nucl-ex]}}.
%%CITATION = ARXIV:1506.08453;%%.

\bibitem{NucleiCERNPS}
V.~T. Cocconi, T.~Fazzini, G.~Fidecaro, M.~Legros, N.~H. Lipman, and A.~W.
  Merrison, ``{Mass Analysis of the Secondary Particles Produced by the 25-Gev
  Proton Beam of the Cern Proton Synchrotron}'',
  \href{http://dx.doi.org/10.1103/PhysRevLett.5.19}{{\em Phys. Rev. Lett.}
  {\bfseries 5} (1960) 19--21}.

\bibitem{ReviewAGS}
S.~Nagamiya, ``{Experimental overview}'',
  \href{http://dx.doi.org/10.1016/0375-9474(92)90562-X}{{\em Nucl. Phys. A}
  {\bfseries 544} (1992) 5C--26C}.

\bibitem{RHIC1}
{\bfseries STAR} Collaboration, C.~Adler {\em et~al.}, ``{Anti-deuteron and
  anti-He-3 production in s(NN)**(1/2) = 130-GeV Au+Au collisions}'',
  \href{http://dx.doi.org/10.1103/PhysRevLett.87.262301}{{\em Phys. Rev. Lett.}
  {\bfseries 87} (2001) 262301},
  \href{http://arxiv.org/abs/nucl-ex/0108022}{{\ttfamily
  arXiv:nucl-ex/0108022}}. [Erratum: Phys.Rev.Lett. 87, 279902 (2001)].

\bibitem{RHIC2}
{\bfseries PHENIX} Collaboration, S.~S. Adler {\em et~al.}, ``{Deuteron and
  antideuteron production in Au + Au collisions at s(NN)**(1/2) = 200-GeV}'',
  \href{http://dx.doi.org/10.1103/PhysRevLett.94.122302}{{\em Phys. Rev. Lett.}
  {\bfseries 94} (2005) 122302},
  \href{http://arxiv.org/abs/nucl-ex/0406004}{{\ttfamily
  arXiv:nucl-ex/0406004}}.

\bibitem{RHIC3}
{\bfseries BRAHMS} Collaboration, I.~Arsene {\em et~al.}, ``{Rapidity
  dependence of deuteron production in Au+Au collisions at $\sqrt{s_{NN}}$ =
  200 GeV}'', \href{http://dx.doi.org/10.1103/PhysRevC.83.044906}{{\em Phys.
  Rev. C} {\bfseries 83} (2011) 044906},
  \href{http://arxiv.org/abs/1005.5427}{{\ttfamily arXiv:1005.5427 [nucl-ex]}}.

\bibitem{RHIC4}
{\bfseries STAR} Collaboration, N.~Yu, ``{Beam energy dependence of $d$ and
  $\bar{d}$ productions in Au+Au collisions at RHIC}'',
  \href{http://dx.doi.org/10.1016/j.nuclphysa.2017.06.046}{{\em Nucl. Phys. A}
  {\bfseries 967} (2017) 788--791},
  \href{http://arxiv.org/abs/1704.04335}{{\ttfamily arXiv:1704.04335
  [nucl-ex]}}.

\bibitem{RHIC5}
{\bfseries STAR} Collaboration, B.~I. Abelev {\em et~al.}, ``{Observation of an
  Antimatter Hypernucleus}'',
  \href{http://dx.doi.org/10.1126/science.1183980}{{\em Science} {\bfseries
  328} (2010) 58--62}, \href{http://arxiv.org/abs/1003.2030}{{\ttfamily
  arXiv:1003.2030 [nucl-ex]}}.

\bibitem{RHIC6}
{\bfseries STAR} Collaboration, H.~Agakishiev {\em et~al.}, ``{Observation of
  the antimatter helium-4 nucleus}'',
  \href{http://dx.doi.org/10.1038/nature10079}{{\em Nature} {\bfseries 473}
  (2011) 353}, \href{http://arxiv.org/abs/1103.3312}{{\ttfamily arXiv:1103.3312
  [nucl-ex]}}. [Erratum: Nature 475, 412 (2011)].

\bibitem{nuclei_pp_PbPb}
{\bfseries ALICE} Collaboration, J.~Adam {\em et~al.}, ``{Production of light
  nuclei and anti-nuclei in pp and Pb-Pb collisions at energies available at
  the CERN Large Hadron Collider}'',
  \href{http://dx.doi.org/10.1103/PhysRevC.93.024917}{{\em Phys. Rev. C}
  {\bfseries 93} no.~2, (2016) 024917},
  \href{http://arxiv.org/abs/1506.08951}{{\ttfamily arXiv:1506.08951
  [nucl-ex]}}.

\bibitem{deuteron_pp_7TeV}
{\bfseries ALICE} Collaboration, S.~Acharya {\em et~al.}, ``{Multiplicity
  dependence of (anti-)deuteron production in pp collisions at $\sqrt{s}$ = 7
  TeV}'', \href{http://dx.doi.org/10.1016/j.physletb.2019.05.028}{{\em Phys.
  Lett.} {\bfseries B794} (2019) 50--63},
\href{http://arxiv.org/abs/1902.09290}{{\ttfamily arXiv:1902.09290 [nucl-ex]}}.
%%CITATION = ARXIV:1902.09290;%%.

\bibitem{deuteron_PbPb_276TeV}
{\bfseries ALICE} Collaboration, S.~Acharya {\em et~al.}, ``{Measurement of
  deuteron spectra and elliptic flow in Pb-Pb collisions at $\sqrt{s_{\mathrm
  {NN}}}$ = 2.76 TeV at the LHC}'',
  \href{http://dx.doi.org/10.1140/epjc/s10052-017-5222-x}{{\em Eur. Phys. J.}
  {\bfseries C77} no.~10, (2017) 658},
\href{http://arxiv.org/abs/1707.07304}{{\ttfamily arXiv:1707.07304 [nucl-ex]}}.
%%CITATION = ARXIV:1707.07304;%%.

\bibitem{nuclei_pp}
{\bfseries ALICE} Collaboration, S.~Acharya {\em et~al.}, ``{Production of
  deuterons, tritons, $^{3}$He nuclei and their antinuclei in pp collisions at
  $\mathbf{\sqrt{{\textit s}}}$ = 0.9, 2.76 and 7 TeV}'',
  \href{http://dx.doi.org/10.1103/PhysRevC.97.024615}{{\em Phys. Rev.}
  {\bfseries C97} no.~2, (2018) 024615},
\href{http://arxiv.org/abs/1709.08522}{{\ttfamily arXiv:1709.08522 [nucl-ex]}}.
%%CITATION = ARXIV:1709.08522;%%.

\bibitem{4He_PbPb}
{\bfseries ALICE} Collaboration, S.~Acharya {\em et~al.}, ``{Production of
  $^{4}$He and $^{4}\overline{\textrm{He}}$ in Pb-Pb collisions at
  $\sqrt{s_{\mathrm{NN}}}$ = 2.76 TeV at the LHC}'',
  \href{http://dx.doi.org/10.1016/j.nuclphysa.2017.12.004}{{\em Nucl. Phys.}
  {\bfseries A971} (2018) 1--20},
\href{http://arxiv.org/abs/1710.07531}{{\ttfamily arXiv:1710.07531 [nucl-ex]}}.
%%CITATION = ARXIV:1710.07531;%%.

\bibitem{deuteron_pPbALICE}
{\bfseries ALICE} Collaboration, S.~Acharya {\em et~al.}, ``{Multiplicity
  dependence of light (anti-)nuclei production in p-Pb collisions at
  $\sqrt{s_{\rm{NN}}}$ = 5.02 TeV}'',
  \href{http://dx.doi.org/10.1016/j.physletb.2019.135043}{{\em Phys. Lett. B}
  {\bfseries 800} (2020) 135043},
  \href{http://arxiv.org/abs/1906.03136}{{\ttfamily arXiv:1906.03136
  [nucl-ex]}}.

\bibitem{3He_pPb}
{\bfseries ALICE} Collaboration, S.~Acharya {\em et~al.}, ``{Production of
  (anti-)$^3$He and (anti-)$^3$H in p-Pb collisions at $\sqrt{s_{\rm{NN}}}$ =
  5.02 TeV}'', \href{http://dx.doi.org/10.1103/PhysRevC.101.044906}{{\em Phys.
  Rev.} {\bfseries C101} no.~4, (2020) 044906},
\href{http://arxiv.org/abs/1910.14401}{{\ttfamily arXiv:1910.14401 [nucl-ex]}}.
%%CITATION = ARXIV:1910.14401;%%.

\bibitem{deuteron_pp_13TeV}
{\bfseries ALICE} Collaboration, S.~Acharya {\em et~al.}, ``{(Anti-)deuteron
  production in pp collisions at $\sqrt{s}=13 \ \text {TeV}$}'',
  \href{http://dx.doi.org/10.1140/epjc/s10052-020-8256-4}{{\em Eur. Phys. J. C}
  {\bfseries 80} no.~9, (2020) 889},
  \href{http://arxiv.org/abs/2003.03184}{{\ttfamily arXiv:2003.03184
  [nucl-ex]}}.

\bibitem{dark_matter1}
K.~Blum, K.~C.~Y. Ng, R.~Sato, and M.~Takimoto, ``{Cosmic rays, antihelium, and
  an old navy spotlight}'',
  \href{http://dx.doi.org/10.1103/PhysRevD.96.103021}{{\em Phys. Rev. D}
  {\bfseries 96} no.~10, (2017) 103021},
  \href{http://arxiv.org/abs/1704.05431}{{\ttfamily arXiv:1704.05431
  [astro-ph.HE]}}.

\bibitem{dark_matter2}
V.~Poulin, P.~Salati, I.~Cholis, M.~Kamionkowski, and J.~Silk, ``{Where do the
  AMS-02 antihelium events come from?}'',
  \href{http://dx.doi.org/10.1103/PhysRevD.99.023016}{{\em Phys. Rev. D}
  {\bfseries 99} no.~2, (2019) 023016},
  \href{http://arxiv.org/abs/1808.08961}{{\ttfamily arXiv:1808.08961
  [astro-ph.HE]}}.

\bibitem{dark_matter3}
M.~Korsmeier, F.~Donato, and N.~Fornengo, ``{Prospects to verify a possible
  dark matter hint in cosmic antiprotons with antideuterons and antihelium}'',
  \href{http://dx.doi.org/10.1103/PhysRevD.97.103011}{{\em Phys. Rev. D}
  {\bfseries 97} no.~10, 103011},
  \href{http://arxiv.org/abs/1711.08465}{{\ttfamily arXiv:1711.08465
  [astro-ph.HE]}}.

\bibitem{dark_matter4}
Y.~Cui, J.~D. Mason, and L.~Randall, ``{General Analysis of Antideuteron
  Searches for Dark Matter}'',
  \href{http://dx.doi.org/10.1007/JHEP11(2010)017}{{\em JHEP} {\bfseries 11}
  (2010) 017}, \href{http://arxiv.org/abs/1006.0983}{{\ttfamily arXiv:1006.0983
  [hep-ph]}}.

\bibitem{SHM1}
J.~Cleymans, S.~Kabana, I.~Kraus, H.~Oeschler, K.~Redlich, and N.~Sharma,
  ``{Antimatter production in proton-proton and heavy-ion collisions at
  ultrarelativistic energies}'',
  \href{http://dx.doi.org/10.1103/PhysRevC.84.054916}{{\em Phys. Rev.}
  {\bfseries C84} (2011) 054916},
\href{http://arxiv.org/abs/1105.3719}{{\ttfamily arXiv:1105.3719 [hep-ph]}}.
%%CITATION = ARXIV:1105.3719;%%.

\bibitem{SHM2}
A.~Andronic, P.~Braun-Munzinger, J.~Stachel, and H.~St$\ddot{\mathrm{o}}$cker,
  ``{Production of light nuclei, hypernuclei and their antiparticles in
  relativistic nuclear collisions}'',
  \href{http://dx.doi.org/10.1016/j.physletb.2011.01.053}{{\em Phys. Lett.}
  {\bfseries B697} (2011) 203--207},
\href{http://arxiv.org/abs/1010.2995}{{\ttfamily arXiv:1010.2995 [nucl-th]}}.
%%CITATION = ARXIV:1010.2995;%%.

\bibitem{SHM3}
F.~Becattini, E.~Grossi, M.~Bleicher, J.~Steinheimer, and R.~Stock,
  ``{Centrality dependence of hadronization and chemical freeze-out conditions
  in heavy ion collisions at $\sqrt{ s_{NN}}$ = 2.76 TeV}'',
  \href{http://dx.doi.org/10.1103/PhysRevC.90.054907}{{\em Phys. Rev.}
  {\bfseries C90} (2014) 054907},
\href{http://arxiv.org/abs/1405.0710}{{\ttfamily arXiv:1405.0710 [nucl-th]}}.
%%CITATION = ARXIV:1405.0710;%%.

\bibitem{SHM4}
V.~Vovchenko and H.~St$\ddot{\mathrm{o}}$cker, ``{Examination of the
  sensitivity of the thermal fits to heavy-ion hadron yield data to the
  modeling of the eigenvolume interactions}'',
  \href{http://dx.doi.org/10.1103/PhysRevC.95.044904}{{\em Phys. Rev.}
  {\bfseries C95} (2017) 044904},
\href{http://arxiv.org/abs/1606.06218}{{\ttfamily arXiv:1606.06218 [hep-ph]}}.
%%CITATION = ARXIV:1606.06218;%%.

\bibitem{SHM5}
A.~Andronic, P.~Braun-Munzinger, K.~Redlich, and J.~Stachel, ``{Decoding the
  phase structure of QCD via particle production at high energy}'',
  \href{http://dx.doi.org/10.1038/s41586-018-0491-6}{{\em Nature} {\bfseries
  561} (2018) 321--330},
\href{http://arxiv.org/abs/1710.09425}{{\ttfamily arXiv:1710.09425 [nucl-th]}}.
%%CITATION = ARXIV:1710.09425;%%.

\bibitem{SHM6}
N.~Sharma, J.~Cleymans, B.~Hippolyte, and M.~Paradza, ``{A Comparison of p-p,
  p-Pb, Pb-Pb Collisions in the Thermal Model: Multiplicity Dependence of
  Thermal Parameters}'',
  \href{http://dx.doi.org/10.1103/PhysRevC.99.044914}{{\em Phys. Rev.}
  {\bfseries C99} (2019) 044914},
\href{http://arxiv.org/abs/1811.00399}{{\ttfamily arXiv:1811.00399 [hep-ph]}}.
%%CITATION = ARXIV:1811.00399;%%.

\bibitem{vanillaCSM}
V.~Vovchenko, B.~D\"onigus, and H.~Stoecker, ``{Canonical statistical model
  analysis of p-p , p -Pb, and Pb-Pb collisions at energies available at the
  CERN Large Hadron Collider}'',
  \href{http://dx.doi.org/10.1103/PhysRevC.100.054906}{{\em Phys. Rev. C}
  {\bfseries 100} no.~5, (2019) 054906},
  \href{http://arxiv.org/abs/1906.03145}{{\ttfamily arXiv:1906.03145
  [hep-ph]}}.

\bibitem{Coalescence1}
S.~T. Butler and C.~A. Pearson, ``{Deuterons from High-Energy Proton
  Bombardment of Matter}'',
\href{http://dx.doi.org/10.1103/PhysRev.129.836}{{\em Phys. Rev.} {\bfseries
  129} (1963) 836--842}.
%%CITATION = PHRVA,129,836;%%.

\bibitem{Coalescence2}
J.~I. Kapusta, ``{Mechanisms for deuteron production in relativistic nuclear
  collisions}'', \href{http://dx.doi.org/10.1103/PhysRevC.21.1301}{{\em Phys.
  Rev. C} {\bfseries 21} (1980) 1301--1310}.

\bibitem{iEBE_VISHNU}
W.~Zhao, L.~Zhu, H.~Zheng, C.~M. Ko, and H.~Song, ``{Spectra and flow of light
  nuclei in relativistic heavy ion collisions at energies available at the BNL
  Relativistic Heavy Ion Collider and at the CERN Large Hadron Collider}'',
  \href{http://dx.doi.org/10.1103/PhysRevC.98.054905}{{\em Phys. Rev.}
  {\bfseries C98} (2018) 054905},
\href{http://arxiv.org/abs/1807.02813}{{\ttfamily arXiv:1807.02813 [nucl-th]}}.
%%CITATION = ARXIV:1807.02813;%%.

\bibitem{Coalescence3}
R.~Scheibl and U.~W. Heinz, ``{Coalescence and flow in ultrarelativistic heavy
  ion collisions}'', \href{http://dx.doi.org/10.1103/PhysRevC.59.1585}{{\em
  Phys. Rev.} {\bfseries C59} (1999) 1585--1602},
\href{http://arxiv.org/abs/nucl-th/9809092}{{\ttfamily arXiv:nucl-th/9809092
  [nucl-th]}}.
%%CITATION = NUCL-TH/9809092;%%.

\bibitem{coalescenceSmallSystems}
K.-J. Sun, C.~M. Ko, and B.~D$\ddot{\mathrm{o}}$nigus, ``{Suppression of light
  nuclei production in collisions of small systems at the Large Hadron
  Collider}'', \href{http://dx.doi.org/10.1016/j.physletb.2019.03.033}{{\em
  Phys. Lett.} {\bfseries B792} (2019) 132--137},
\href{http://arxiv.org/abs/1812.05175}{{\ttfamily arXiv:1812.05175 [nucl-th]}}.
%%CITATION = ARXIV:1812.05175;%%.

\bibitem{alternativeCoalescence}
M.~Kachelrie{\ss}, S.~Ostapchenko, and J.~Tjemsland, ``{Alternative coalescence
  model for deuteron, tritium, helium-3 and their antinuclei}'',
  \href{http://dx.doi.org/10.1140/epja/s10050-019-00007-9}{{\em Eur. Phys. J.
  A} {\bfseries 56} no.~1, (2020) 4},
  \href{http://arxiv.org/abs/1905.01192}{{\ttfamily arXiv:1905.01192
  [hep-ph]}}.

\bibitem{DeuteronInJets}
{\bfseries ALICE} Collaboration, S.~Acharya {\em et~al.}, ``{Jet-associated
  deuteron production in pp collisions at $\sqrt{s}$=13 TeV}'',
  \href{http://dx.doi.org/https://doi.org/10.1016/j.physletb.2021.136440}{{\em
  Physics Letters B} {\bfseries 819} (2021) 136440}.
  \url{https://www.sciencedirect.com/science/article/pii/S0370269321003804}.

\bibitem{coalescence_correlations}
K.~Blum and M.~Takimoto, ``{Nuclear coalescence from correlation functions}'',
  \href{http://dx.doi.org/10.1103/PhysRevC.99.044913}{{\em Phys. Rev. C}
  {\bfseries 99} no.~4, (2019) 044913},
  \href{http://arxiv.org/abs/1901.07088}{{\ttfamily arXiv:1901.07088
  [nucl-th]}}.

\bibitem{sourceSizeHMpp}
{\bfseries ALICE} Collaboration, S.~Acharya {\em et~al.}, ``{Search for a
  common baryon source in high-multiplicity pp collisions at the LHC}'',
  \href{http://dx.doi.org/10.1016/j.physletb.2020.135849}{{\em Phys. Lett. B}
  {\bfseries 811} (2020) 135849},
  \href{http://arxiv.org/abs/2004.08018}{{\ttfamily arXiv:2004.08018
  [nucl-ex]}}.

\bibitem{ALICEexperiment}
{\bfseries ALICE} Collaboration, K.~Aamodt {\em et~al.}, ``{The ALICE
  experiment at the CERN LHC}'',
\href{http://dx.doi.org/10.1088/1748-0221/3/08/S08002}{{\em JINST} {\bfseries
  3} (2008) S08002}.
%%CITATION = JINST,3,S08002;%%.

\bibitem{ALICEperformance}
{\bfseries ALICE} Collaboration, B.~Abelev {\em et~al.}, ``{Performance of the
  ALICE Experiment at the CERN LHC}'',
  \href{http://dx.doi.org/10.1142/S0217751X14300440}{{\em Int. J. Mod. Phys.}
  {\bfseries A29} (2014) 1430044},
\href{http://arxiv.org/abs/1402.4476}{{\ttfamily arXiv:1402.4476 [nucl-ex]}}.
%%CITATION = ARXIV:1402.4476;%%.

\bibitem{ITS}
{\bfseries ALICE} Collaboration, K.~Aamodt {\em et~al.}, ``{Alignment of the
  ALICE Inner Tracking System with cosmic-ray tracks}'',
  \href{http://dx.doi.org/10.1088/1748-0221/5/03/P03003}{{\em JINST} {\bfseries
  5} (2010) P03003},
\href{http://arxiv.org/abs/1001.0502}{{\ttfamily arXiv:1001.0502
  [physics.ins-det]}}.
%%CITATION = ARXIV:1001.0502;%%.

\bibitem{TPC}
J.~Alme {\em et~al.}, ``{The ALICE TPC, a large 3-dimensional tracking device
  with fast readout for ultra-high multiplicity events}'',
  \href{http://dx.doi.org/10.1016/j.nima.2010.04.042}{{\em Nucl. Instrum.
  Meth.} {\bfseries A622} (2010) 316--367},
\href{http://arxiv.org/abs/1001.1950}{{\ttfamily arXiv:1001.1950
  [physics.ins-det]}}.
%%CITATION = ARXIV:1001.1950;%%.

\bibitem{TOF}
{\bfseries ALICE} Collaboration, A.~Akindinov {\em et~al.}, ``{Performance of
  the ALICE Time-Of-Flight detector at the LHC}'',
\href{http://dx.doi.org/10.1140/epjp/i2013-13044-x}{{\em Eur. Phys. J. Plus}
  {\bfseries 128} (2013) 44}.
%%CITATION = EPHJP,128,44;%%.

\bibitem{VZEROPerformance}
{\bfseries ALICE} Collaboration, E.~Abbas {\em et~al.}, ``{Performance of the
  ALICE VZERO system}'',
  \href{http://dx.doi.org/10.1088/1748-0221/8/10/P10016}{{\em JINST} {\bfseries
  8} (2013) P10016},
\href{http://arxiv.org/abs/1306.3130}{{\ttfamily arXiv:1306.3130 [nucl-ex]}}.
%%CITATION = ARXIV:1306.3130;%%.

\bibitem{mutliplicity7tev}
{\bfseries ALICE} Collaboration, S.~Acharya {\em et~al.}, ``{Multiplicity
  dependence of light-flavor hadron production in pp collisions at $\sqrt{s}$ =
  7 TeV}'', \href{http://dx.doi.org/10.1103/PhysRevC.99.024906}{{\em Phys. Rev.
  C} {\bfseries 99} no.~2, (2019) 024906},
  \href{http://arxiv.org/abs/1807.11321}{{\ttfamily arXiv:1807.11321
  [nucl-ex]}}.

\bibitem{strangeness_vs_mult}
{\bfseries ALICE} Collaboration, S.~Acharya {\em et~al.}, ``{Multiplicity
  dependence of (multi-)strange hadron production in proton-proton collisions
  at $\sqrt{s}$ = 13 TeV}'',
  \href{http://dx.doi.org/10.1140/epjc/s10052-020-7673-8}{{\em Eur. Phys. J. C}
  {\bfseries 80} no.~2, (2020) 167},
  \href{http://arxiv.org/abs/1908.01861}{{\ttfamily arXiv:1908.01861
  [nucl-ex]}}.

\bibitem{multiplicity_measurement}
{\bfseries ALICE} Collaboration, B.~Abelev {\em et~al.}, ``{Pseudorapidity
  density of charged particles in $p$ + Pb collisions at $\sqrt{s_{NN}}=5.02$
  TeV}'', \href{http://dx.doi.org/10.1103/PhysRevLett.110.032301}{{\em Phys.
  Rev. Lett.} {\bfseries 110} no.~3, (2013) 032301},
  \href{http://arxiv.org/abs/1210.3615}{{\ttfamily arXiv:1210.3615 [nucl-ex]}}.

\bibitem{Pythia8}
T.~Sjostrand, S.~Mrenna, and P.~Z. Skands, ``{A Brief Introduction to PYTHIA
  8.1}'', \href{http://dx.doi.org/10.1016/j.cpc.2008.01.036}{{\em Comput. Phys.
  Commun.} {\bfseries 178} (2008) 852--867},
\href{http://arxiv.org/abs/0710.3820}{{\ttfamily arXiv:0710.3820 [hep-ph]}}.
%%CITATION = ARXIV:0710.3820;%%.

\bibitem{Pythia8Monash2013}
P.~Skands, S.~Carrazza, and J.~Rojo, ``{Tuning PYTHIA 8.1: the Monash 2013
  Tune}'', \href{http://dx.doi.org/10.1140/epjc/s10052-014-3024-y}{{\em Eur.
  Phys. J. C} {\bfseries 74} no.~8, (2014) 3024},
  \href{http://arxiv.org/abs/1404.5630}{{\ttfamily arXiv:1404.5630 [hep-ph]}}.

\bibitem{GEANT4}
{\bfseries GEANT4} Collaboration, S.~Agostinelli {\em et~al.}, ``{GEANT4--a
  simulation toolkit}'',
  \href{http://dx.doi.org/10.1016/S0168-9002(03)01368-8}{{\em Nucl. Instrum.
  Meth. A} {\bfseries 506} (2003) 250--303}.

\bibitem{antideuteronInelCS}
{\bfseries ALICE} Collaboration, S.~Acharya {\em et~al.}, ``{Measurement of the
  low-energy antideuteron inelastic cross section}'',
  \href{http://dx.doi.org/10.1103/PhysRevLett.125.162001}{{\em Phys. Rev.
  Lett.} {\bfseries 125} no.~16, (2020) 162001}.

\bibitem{GEANT3}
R.~Brun, F.~Bruyant, F.~Carminati, S.~Giani, M.~Maire, A.~McPherson,
  G.~Patrick, and L.~Urban, ``{GEANT Detector Description and Simulation
  Tool}'', \href{http://dx.doi.org/10.17181/CERN.MUHF.DMJ1}{ (1994)
  CERN--W5013}.

\bibitem{Tsallis}
C.~Tsallis, ``Possible generalization of {B}oltzmann-{G}ibbs statistics'',
  \href{http://dx.doi.org/10.1007/BF01016429}{{\em Journal of statistical
  physics} {\bfseries 52} no.~1-2, (1988) 479--487}.

\bibitem{BlastWave1}
E.~Schnedermann, J.~Sollfrank, and U.~W. Heinz, ``{Thermal phenomenology of
  hadrons from 200-A/GeV S+S collisions}'',
  \href{http://dx.doi.org/10.1103/PhysRevC.48.2462}{{\em Phys. Rev.} {\bfseries
  C48} (1993) 2462--2475},
\href{http://arxiv.org/abs/nucl-th/9307020}{{\ttfamily arXiv:nucl-th/9307020
  [nucl-th]}}.
%%CITATION = NUCL-TH/9307020;%%.

\bibitem{BlastWave2}
{\bfseries STAR} Collaboration, C.~Adler {\em et~al.}, ``{Identified particle
  elliptic flow in Au + Au collisions at $\sqrt{s_{\mathrm{NN}}}$ = 130 GeV}'',
  \href{http://dx.doi.org/10.1103/PhysRevLett.87.182301}{{\em Phys. Rev. Lett.}
  {\bfseries 87} (2001) 182301},
\href{http://arxiv.org/abs/nucl-ex/0107003}{{\ttfamily arXiv:nucl-ex/0107003
  [nucl-ex]}}.
%%CITATION = NUCL-EX/0107003;%%.

\bibitem{BlastWave3}
P.~J. Siemens and J.~O. Rasmussen, ``Evidence for a blast wave from compressed
  nuclear matter'', \href{http://dx.doi.org/10.1103/PhysRevLett.42.880}{{\em
  Phys. Rev. Lett.} {\bfseries 42} (1979) 880--883}.

\bibitem{coalescenceFemto}
F.~Bellini, K.~Blum, A.~P. Kalweit, and M.~Puccio, ``{Examination of
  coalescence as the origin of nuclei in hadronic collisions}'',
  \href{http://dx.doi.org/10.1103/PhysRevC.103.014907}{{\em Phys. Rev. C}
  {\bfseries 103} no.~1, (2021) 014907},
  \href{http://arxiv.org/abs/2007.01750}{{\ttfamily arXiv:2007.01750
  [nucl-th]}}.

\bibitem{chiEFT}
D.~R. Entem, R.~Machleidt, and Y.~Nosyk, ``{High-quality two-nucleon potentials
  up to fifth order of the chiral expansion}'',
  \href{http://dx.doi.org/10.1103/PhysRevC.96.024004}{{\em Phys. Rev. C}
  {\bfseries 96} no.~2, (2017) 024004},
  \href{http://arxiv.org/abs/1703.05454}{{\ttfamily arXiv:1703.05454
  [nucl-th]}}.

\bibitem{coalescenceBelliniKalweit}
F.~Bellini and A.~P. Kalweit, ``Testing production scenarios for
  (anti-)(hyper-)nuclei and exotica at energies available at the cern large
  hadron collider'', \href{http://dx.doi.org/10.1103/PhysRevC.99.054905}{{\em
  Phys. Rev. C} {\bfseries 99} (May, 2019) 054905}.
  \url{https://link.aps.org/doi/10.1103/PhysRevC.99.054905}.

\bibitem{Phillips_1959}
R.~J.~N. Phillips, ``The two-nucleon interaction'',
  \href{http://dx.doi.org/10.1088/0034-4885/22/1/314}{{\em Reports on Progress
  in Physics} {\bfseries 22} no.~1, (Jan, 1959) 562--634}.
  \url{https://doi.org/10.1088/0034-4885/22/1/314}.

\bibitem{massDifference}
E.~Tiesinga, P.~J. Mohr, D.~B. Newell, and B.~N. Taylor, ``Codata recommended
  values of the fundamental physical constants: 2018'',
  \href{http://dx.doi.org/10.1103/RevModPhys.93.025010}{{\em Rev. Mod. Phys.}
  {\bfseries 93} (Jun, 2021) 025010}.

\bibitem{CSM}
V.~Vovchenko, B.~D{\"o}nigus, and H.~Stoecker, ``{Multiplicity dependence of
  light nuclei production at LHC energies in the canonical statistical
  model}'', \href{http://dx.doi.org/10.1016/j.physletb.2018.08.041}{{\em Phys.
  Lett.} {\bfseries B785} (2018) 171--174},
\href{http://arxiv.org/abs/1808.05245}{{\ttfamily arXiv:1808.05245 [hep-ph]}}.
%%CITATION = ARXIV:1808.05245;%%.

\bibitem{pnRun3}
{\bfseries ALICE} Collaboration, S.~Acharya {\em et~al.}, ``{Future high-energy
  pp programme with ALICE}'',. \url{https://cds.cern.ch/record/2724925}.
  ALICE-PUBLIC-2020-005.

\bibitem{ITSU}
{\bfseries ALICE} Collaboration, B.~Abelev {\em et~al.}, ``{Technical Design
  Report for the Upgrade of the ALICE Inner Tracking System}'',
  \href{http://dx.doi.org/10.1088/0954-3899/41/8/087002}{{\em J. Phys. G}
  {\bfseries 41} (2014) 087002}.

\bibitem{oldHBTradii}
{\bfseries ALICE} Collaboration, B.~Abelev {\em et~al.}, ``{Charged kaon
  femtoscopic correlations in $pp$ collisions at $\sqrt{s}=7$ TeV}'',
  \href{http://dx.doi.org/10.1103/PhysRevD.87.052016}{{\em Phys. Rev. D}
  {\bfseries 87} no.~5, (2013) 052016},
  \href{http://arxiv.org/abs/1212.5958}{{\ttfamily arXiv:1212.5958 [hep-ex]}}.

\bibitem{chiEFTnorm}
R.~Machleidt, ``{The High precision, charge dependent Bonn nucleon-nucleon
  potential (CD-Bonn)}'',
  \href{http://dx.doi.org/10.1103/PhysRevC.63.024001}{{\em Phys. Rev. C}
  {\bfseries 63} (2001) 024001},
  \href{http://arxiv.org/abs/nucl-th/0006014}{{\ttfamily
  arXiv:nucl-th/0006014}}.

\end{thebibliography}\endgroup

%%%%%%%%%%%%%%%%%%%%%%%%%%%%%%%%
% Appendices: yours (if any) + authorlist
%%%%%%%%%%%%%%%%%%%%%%%%%%%%%%%%
\newpage
\appendix

\section{Theoretical prediction for the coalescence parameter \texorpdfstring{$B_A$}{BA}}
\label{app:theory}
In this appendix, the details about the theoretical prediction used for the coalescence parameter $B_A$ as a function of the source radius are reported. The general recipe is taken from Ref.~\cite{coalescence_correlations}. $B_2$ is defined as
\begin{equation}
  B_{2} (p)\approx \frac{3}{2m}\int \mathrm{d}^3q \, D(\vec{q}) \, \mathcal{C}_2^{\mathrm{PRF}}\left(\vec{p},\vec{q}\right),
  \label{eq:recipe}
\end{equation}
where $m$ is the proton mass, $p$ is the momentum of the nucleus, $q$ is the relative momentum of the nucleons, $D(\vec{q})$ is the deuteron Wigner density and $\mathcal{C}_2^{\mathrm{PRF}}\left(\vec{p},\vec{q}\right)$ is the correlation between two nucleons in the rest frame of the pair (PRF), assuming a Gaussian source model. For these calculations, we assume a homogeneous source, i.e. $R = R_{\parallel} = R_{\perp}$. Hence, the correlation function has the form
\begin{equation}
  \mathcal{C}_2^{\mathrm{PRF}}\left(\vec{p},\vec{q}\right) = e^{-R^2q^2},
  \label{eq:correlation}
\end{equation}
where $R$ is the source radius. Finally, the Wigner density is defined as
\begin{equation}
  D(\vec{q}) = \int \mathrm{d}^3r \, |\phi_{\mathrm{d}}(\vec{r})|^2 e^{-i \vec{q}\times\vec{r}},
  \label{eq:wigner}
\end{equation}
where $\phi_{\mathrm{d}}$ is the deuteron wave function. In the following, we will provide different predictions for $B_2$ as a function of the source radius $R$ starting from Eq.~\ref{eq:recipe} and using different wave functions $\phi_{\mathrm{d}}$. The theoretical predictions for $B_2$ as a function of the source radius $R$ is shown in Fig.~\ref{fig:BA}(a). At large values of the source radius they all show the same trend. On the contrary, for small values they differ, with a maximum spread of around a factor of 10. Eq.~\ref{eq:recipe} has not an equivalent for $B_3$ and \textit{ab initio} calculations are needed. For this reason, it is currently not possible to obtain $B_3$ predictions for different wave functions as easily as for $B_2$. However, it is possible to obtain a prediction for the case of a simple Gaussian wave function (see Eq.~\ref{eq:BA_gauss}).

\subsection{Gaussian wave function}
The most simple assumption is a Gaussian wave function
\begin{equation}
  \phi_{\mathrm{d}}(r) = \frac{e^{-\frac{r^2}{2d^2}}}{\left(\pi d^2 \right)^{3/4}},
  \label{eq:gaussian}
\end{equation}
where $d$ is the nucleus radius. For this calculations, $d = 3.2$~fm, as in Ref.~\cite{coalescence_correlations}. The corresponding Wigner density is
\begin{equation}
  D(\vec{q}) = e^{-\frac{q^2d^2}{4}}.
\end{equation}
This brings to the expression for $B_2$ as a function of the source radius $R$
\begin{equation}
  B_{2}(R) = \frac{3 \pi^2}{2 m \left[R^2 + \left(\frac{d}{2}\right)^2\right]^{\frac{3}{2}}}.
  \label{eq:B2_gauss}
\end{equation}
This function is shown in Fig.~\ref{fig:BA}(a), together with the other predictions for $B_2$.
As shown in Ref.~\cite{coalescence_correlations}, Eq.~\ref{eq:B2_gauss} can also be generalised for a nucleus with mass number $A$
\begin{equation}
  B_{A}(R) = \frac{2J_{A} + 1}{2^{A}\sqrt{A}}\frac{1}{m^{A-1}}\left[\frac{2\pi}{R^{2} + \left(r_{A}/2\right)^2}\right]^{\frac{3}{2}(A-1)},
  \label{eq:BA_gauss}
\end{equation}
where $J_{A}$ is the spin of the nucleus. Eq.~\ref{eq:BA_gauss} is used to calculate the theoretical prediction for $B_3$, shown in Fig.~\ref{fig:BA}(b).

\subsection{Hulthen wave function}
The second hypothesis tested here is a Hulthen wave function
\begin{equation}
  \phi_{\mathrm{d}}(r) = \sqrt{\frac{\alpha \beta (\alpha + \beta)}{2\pi(\alpha - \beta)^2}} \, \frac{e^{-\alpha r} - e^{-\beta r}}{r},
\end{equation}
where $\alpha = 0.2~\text{fm}^{-1}$ and $\beta = 1.56~\text{fm}^{-1}$ are parameters taken from Ref.~\cite{Coalescence3}. The corresponding expression for $B_2$ is
\begin{equation}
  B_{2}(R) = \frac{3 \pi^2}{R^2} \frac{\alpha\beta(\alpha + \beta)}{(\alpha - \beta)^2} \left[e^{4\alpha^2R^2} \text{erfc}(2\alpha R) - 2 e^{(\alpha+\beta)^2R^2} \text{erfc}((\alpha+\beta) R) + e^{4\beta^2R^2} \text{erfc}(2\beta R)\right].
  \label{eq:B2_hulthen}
\end{equation}
This function is shown in Fig.~\ref{fig:BA}(a), together with the other predictions for $B_2$.

\subsection{Chiral Effective Field Theory wave function}
The third hypothesis for deuteron wave function is obtained from Chiral Effective field theory ($\chi \text{EFT}$) calculations ($\text{N}^4\text{LO}$). It is based on Ref.~\cite{chiEFT} and the normalisation is based on Ref.~\cite{chiEFTnorm}. A cutoff at $\Lambda_c = 500$~MeV is used. The deuteron wave function is
\begin{equation}
  \phi_{\mathrm{d}}(\vec{r}) = \frac{1}{\sqrt{4\pi} \, r}\left[u(r) \, + \, \frac{1}{\sqrt{8}}\,  w(r) \, S_{12}(\hat{r})\right]\, \chi_{1m}\, , 
  \label{eq:deut_wf_CET}
\end{equation}
where $u(r)$ and $w(r)$ are radial wave functions, $S_{12}(\hat{r})$ is the spin tensor and $\chi_{1m}$ is a spinor. The spin-averaged density of the deuteron can be hence expressed as
\begin{equation}
  \left|\phi_{\mathrm{d}}(r)\right|^2 = \frac{1}{4\pi \, r^2}\left[u^2(r) \, + \, w^2(r)\right].
  \label{eq:deut_density_CET}
\end{equation}
Using Eq.~\ref{eq:recipe} and Eq.~\ref{eq:wigner}, one obtains
\begin{equation}
  B_2(R) = \frac{6\pi}{m}\int_{0}^{\Lambda_c}\mathrm{d}q \int_{0}^{\infty}\mathrm{d}r \, q \, \left[ u(r)^2 \, + \, w(r)^2 \right]\,\frac{\sin(qr)}{r} \, e^{-R^2q^2} .
  \label{eq:b2_cet_difficult}
\end{equation}
Integrating Eq.~\ref{eq:b2_cet_difficult} over $q$, one obtains
\begin{dmath}
  B_2(R) = \frac{3 \pi}{m R^2}\int_{0}^{\infty} \,\frac{\mathrm{d}r}{r} \, \left[u^2(r)\, + \, w^2(r)\right] \, \left\{e^{-\Lambda_c^2R^2}\sin(\Lambda_c r) \, + \,\frac{\sqrt{\pi}r} {4R} \, e^{-\frac{r^2}{4R^2}} \, \left[\erf \left(\frac{ir+2R^2\Lambda_c}{2R}\right) \, - \, \erf \left(\frac{ir-2R^2\Lambda_c}{2R}\right)\right]\right\}\, .
  \label{eq:b2_cet_simple}
\end{dmath}
This function is shown in Fig.~\ref{fig:BA}(a), together with the other predictions for $B_2$.

%%%%%%%%%%%%%%%%%%%%%%%%%%%%%%%%%%%%%%%%%%%%%
\subsection{Combination of two Gaussians}
\begin{figure}[ht]
  \centering
  \captionsetup[subfigure]{justification=centering}
  \begin{subfigure}[t]{0.49\textwidth}
    \includegraphics[width=\textwidth]{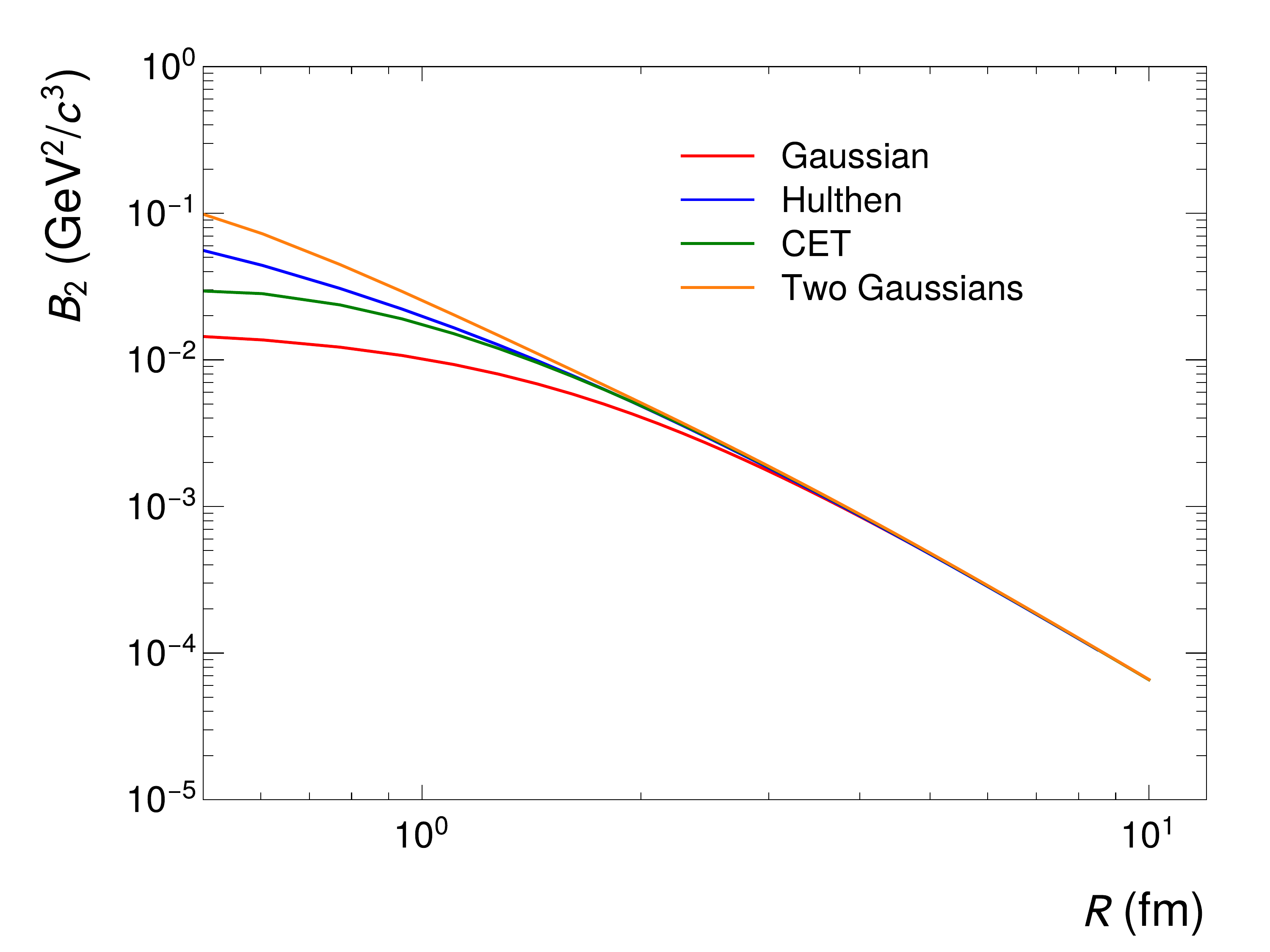}
    \caption{}
    \label{fig:B2}
  \end{subfigure}
  \begin{subfigure}[t]{0.49\textwidth}
    \includegraphics[width=\textwidth]{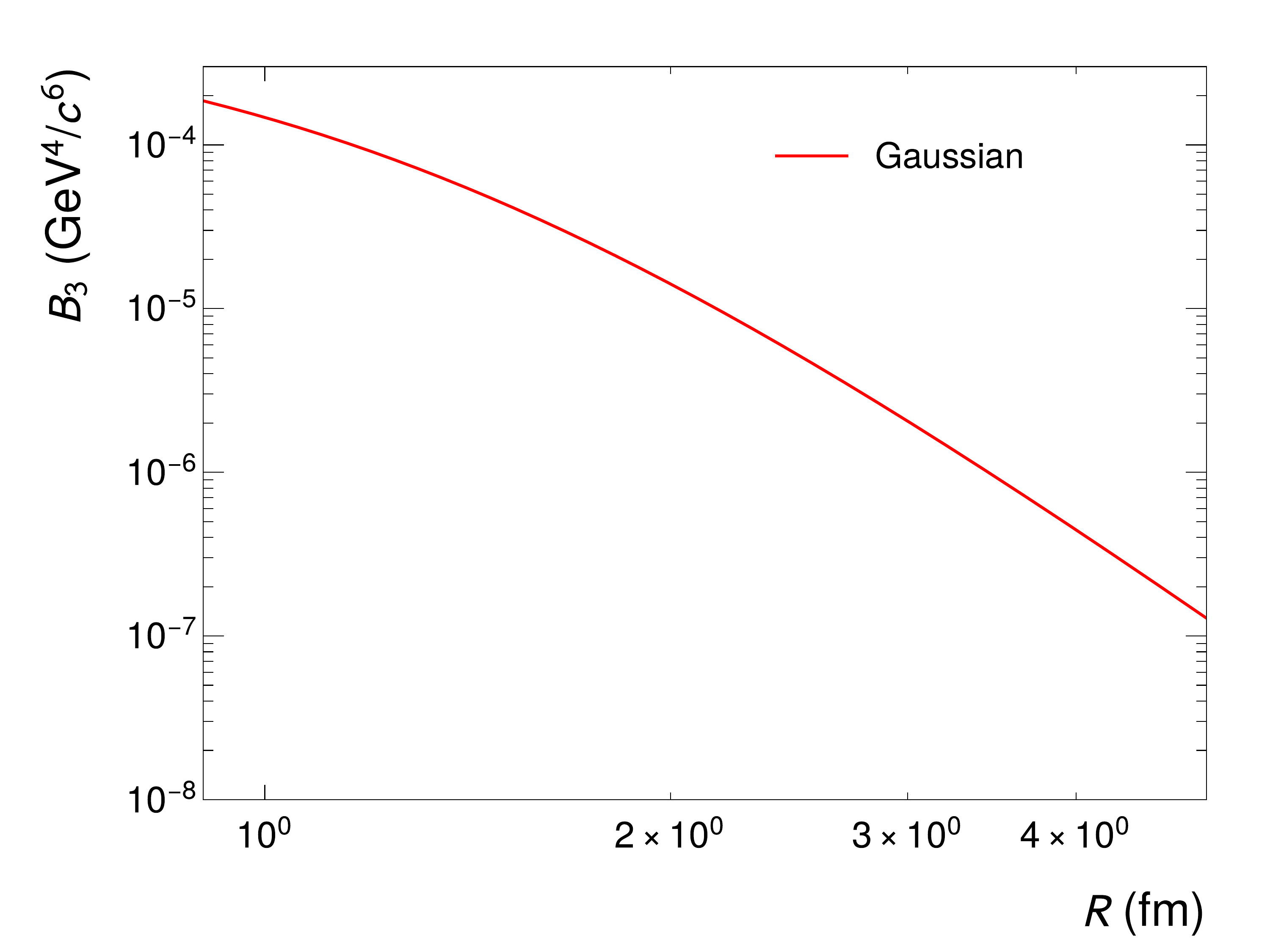}
    \caption{}
    \label{fig:B3}
  \end{subfigure}
  \caption{Coalescence parameters $B_2$ (a) and $B_3$ (b) as a function of the source radius $R$ for different wave functions (see the text for more details).}
  \label{fig:BA}
\end{figure}

The last considered wave function is a combination of two Gaussians, fitted to the Hulthen wave function~\cite{alternativeCoalescence}:
\begin{equation}
  \phi_{\mathrm{d}}(r) = \pi^{-3/4}\left[\frac{\Delta^{1/2}}{d_1^{3/2}}e^{-r^2/(2d_1^2)} \; + \; e^{i\alpha} \frac{(1-\Delta)^{1/2}}{d_2^{3/2}}e^{-r^2/(2d_2^2)}\right]
  \label{eq:two_gauss}
\end{equation} ,
where $\Delta = 0.581$, $d_1 = 3.979$~fm and $d_1 = 0.890$~fm. The corresponding density is
\begin{equation}
|\phi_{\mathrm{d}}(r)|^2 = {\pi^{-3/2}}\left[ \frac{\Delta}{d_1^3}e^{-r^2/d_1^2}+\frac{1-\Delta}{d_2^3}e^{-r^2/d_2^2} \right] \, .
\label{eq:two_gauss_denisty}
\end{equation} 
$B_2$ can be hence written as
\begin{equation}
 B_2(R) = \frac{24 \pi^{5/2}}{m}\int_0^\infty \, \mathrm{d}q \int_0^\infty \, \mathrm{d}r\, |\phi_{d}(r)|^2 \,\sin(qr) \, r \, q \, e^{-R^2q^2}\, ,
 \label{eq:b2_two_gauss_intermatiate_b2} 
\end{equation}
and after integrating over $q$ and $r$, one obtains:
\begin{equation}
  B_2(R) = \frac{3 \pi^{3/2}}{2 m R^3}\left[\Delta \, \left(1 \, + \, \frac{d_1^2}{4R^2}\right)^{-3/2} \, + \, (1 \, - \, \Delta) \, \left(1 \, + \, \frac{d_2^2}{4R^2}\right)^{-3/2}\,\right] \, .
  \label{eq:b2_two_gauss_b2} 
\end{equation}
This function is shown in Fig.~\ref{fig:BA}(a), together with the other predictions for $B_2$.

\newpage

%%%%% Authorlist - please do not touch: handled by EB chairs 
\section{The ALICE Collaboration}
\label{app:collab}
% ALICE Collaboration author list for 2021-08-25

\small
\begin{flushleft} 

S.~Acharya$^{\rm 144}$, 
D.~Adamov\'{a}$^{\rm 98}$, 
A.~Adler$^{\rm 76}$, 
J.~Adolfsson$^{\rm 83}$, 
G.~Aglieri Rinella$^{\rm 35}$, 
M.~Agnello$^{\rm 31}$, 
N.~Agrawal$^{\rm 55}$, 
Z.~Ahammed$^{\rm 144}$, 
S.~Ahmad$^{\rm 16}$, 
S.U.~Ahn$^{\rm 78}$
% ...

\section*{Affiliation Notes}

$^{\rm I}$ Deceased\\
$^{\rm II}$ Also at: Italian National Agency for New Technologies, Energy and Sustainable Economic Development (ENEA), Bologna, Italy\\
$^{\rm III}$ Also at: Dipartimento DET del Politecnico di Torino, Turin, Italy\\
$^{\rm IV}$ Also at: M.V. Lomonosov Moscow State University, D.V. Skobeltsyn Institute of Nuclear, Physics, Moscow, Russia\\
$^{\rm V}$ Also at: Department of Applied Physics, Aligarh Muslim University, Aligarh, India
\\
$^{\rm VI}$ Also at: Institute of Theoretical Physics, University of Wroclaw, Poland\\
$^{\rm VII}$ Also at: University of Kansas, Lawrence, Kansas, United States\\

\section*{Collaboration Institutes}

$^{1}$ A.I. Alikhanyan National Science Laboratory (Yerevan Physics Institute) Foundation, Yerevan, Armenia\\
$^{2}$ AGH University of Science and Technology, Cracow, Poland\\
$^{3}$ Bogolyubov Institute for Theoretical Physics, National Academy of Sciences of Ukraine, Kiev, Ukraine\\
$^{4}$ Bose Institute, Department of Physics  and Centre for Astroparticle Physics and Space Science (CAPSS), Kolkata, India\\
$^{5}$ Budker Institute for Nuclear Physics, Novosibirsk, Russia\\
$^{6}$ California Polytechnic State University, San Luis Obispo, California, United States\\
$^{7}$ Central China Normal University, Wuhan, China\\
$^{8}$ Centro de Aplicaciones Tecnol\'{o}gicas y Desarrollo Nuclear (CEADEN), Havana, Cuba\\
$^{9}$ Centro de Investigaci\'{o}n y de Estudios Avanzados (CINVESTAV), Mexico City and M\'{e}rida, Mexico\\
$^{10}$ Chicago State University, Chicago, Illinois, United States\\
$^{11}$ China Institute of Atomic Energy, Beijing, China\\
$^{12}$ Chungbuk National University, Cheongju, Republic of Korea\\
$^{13}$ Comenius University Bratislava, Faculty of Mathematics, Physics and Informatics, Bratislava, Slovakia\\
$^{14}$ COMSATS University Islamabad, Islamabad, Pakistan\\
$^{15}$ Creighton University, Omaha, Nebraska, United States\\
$^{16}$ Department of Physics, Aligarh Muslim University, Aligarh, India\\
$^{17}$ Department of Physics, Pusan National University, Pusan, Republic of Korea\\
$^{18}$ Department of Physics, Sejong University, Seoul, Republic of Korea\\
$^{19}$ Department of Physics, University of California, Berkeley, California, United States\\
$^{20}$ Department of Physics, University of Oslo, Oslo, Norway\\
$^{21}$ Department of Physics and Technology, University of Bergen, Bergen, Norway\\
$^{22}$ Dipartimento di Fisica dell'Universit\`{a} 'La Sapienza' and Sezione INFN, Rome, Italy\\
$^{23}$ Dipartimento di Fisica dell'Universit\`{a} and Sezione INFN, Cagliari, Italy\\
$^{24}$ Dipartimento di Fisica dell'Universit\`{a} and Sezione INFN, Trieste, Italy\\
$^{25}$ Dipartimento di Fisica dell'Universit\`{a} and Sezione INFN, Turin, Italy\\
$^{26}$ Dipartimento di Fisica e Astronomia dell'Universit\`{a} and Sezione INFN, Bologna, Italy\\
$^{27}$ Dipartimento di Fisica e Astronomia dell'Universit\`{a} and Sezione INFN, Catania, Italy\\
$^{28}$ Dipartimento di Fisica e Astronomia dell'Universit\`{a} and Sezione INFN, Padova, Italy\\
$^{29}$ Dipartimento di Fisica e Nucleare e Teorica, Universit\`{a} di Pavia, Pavia, Italy\\
$^{30}$ Dipartimento di Fisica `E.R.~Caianiello' dell'Universit\`{a} and Gruppo Collegato INFN, Salerno, Italy\\
$^{31}$ Dipartimento DISAT del Politecnico and Sezione INFN, Turin, Italy\\
$^{32}$ Dipartimento di Scienze e Innovazione Tecnologica dell'Universit\`{a} del Piemonte Orientale and INFN Sezione di Torino, Alessandria, Italy\\
$^{33}$ Dipartimento di Scienze MIFT, Universit\`{a} di Messina, Messina, Italy\\
$^{34}$ Dipartimento Interateneo di Fisica `M.~Merlin' and Sezione INFN, Bari, Italy\\
$^{35}$ European Organization for Nuclear Research (CERN), Geneva, Switzerland\\
$^{36}$ Faculty of Electrical Engineering, Mechanical Engineering and Naval Architecture, University of Split, Split, Croatia\\
$^{37}$ Faculty of Engineering and Science, Western Norway University of Applied Sciences, Bergen, Norway\\
$^{38}$ Faculty of Nuclear Sciences and Physical Engineering, Czech Technical University in Prague, Prague, Czech Republic\\
$^{39}$ Faculty of Science, P.J.~\v{S}af\'{a}rik University, Ko\v{s}ice, Slovakia\\
$^{40}$ Frankfurt Institute for Advanced Studies, Johann Wolfgang Goethe-Universit\"{a}t Frankfurt, Frankfurt, Germany\\
$^{41}$ Fudan University, Shanghai, China\\
$^{42}$ Gangneung-Wonju National University, Gangneung, Republic of Korea\\
$^{43}$ Gauhati University, Department of Physics, Guwahati, India\\
$^{44}$ Helmholtz-Institut f\"{u}r Strahlen- und Kernphysik, Rheinische Friedrich-Wilhelms-Universit\"{a}t Bonn, Bonn, Germany\\
$^{45}$ Helsinki Institute of Physics (HIP), Helsinki, Finland\\
$^{46}$ High Energy Physics Group,  Universidad Aut\'{o}noma de Puebla, Puebla, Mexico\\
$^{47}$ Hiroshima University, Hiroshima, Japan\\
$^{48}$ Hochschule Worms, Zentrum  f\"{u}r Technologietransfer und Telekommunikation (ZTT), Worms, Germany\\
$^{49}$ Horia Hulubei National Institute of Physics and Nuclear Engineering, Bucharest, Romania\\
$^{50}$ Indian Institute of Technology Bombay (IIT), Mumbai, India\\
$^{51}$ Indian Institute of Technology Indore, Indore, India\\
$^{52}$ Indonesian Institute of Sciences, Jakarta, Indonesia\\
$^{53}$ INFN, Laboratori Nazionali di Frascati, Frascati, Italy\\
$^{54}$ INFN, Sezione di Bari, Bari, Italy\\
$^{55}$ INFN, Sezione di Bologna, Bologna, Italy\\
$^{56}$ INFN, Sezione di Cagliari, Cagliari, Italy\\
$^{57}$ INFN, Sezione di Catania, Catania, Italy\\
$^{58}$ INFN, Sezione di Padova, Padova, Italy\\
$^{59}$ INFN, Sezione di Pavia, Pavia, Italy\\
$^{60}$ INFN, Sezione di Roma, Rome, Italy\\
$^{61}$ INFN, Sezione di Torino, Turin, Italy\\
$^{62}$ INFN, Sezione di Trieste, Trieste, Italy\\
$^{63}$ Inha University, Incheon, Republic of Korea\\
$^{64}$ Institute for Gravitational and Subatomic Physics (GRASP), Utrecht University/Nikhef, Utrecht, Netherlands\\
$^{65}$ Institute for Nuclear Research, Academy of Sciences, Moscow, Russia\\
$^{66}$ Institute of Experimental Physics, Slovak Academy of Sciences, Ko\v{s}ice, Slovakia\\
$^{67}$ Institute of Physics, Homi Bhabha National Institute, Bhubaneswar, India\\
$^{68}$ Institute of Physics of the Czech Academy of Sciences, Prague, Czech Republic\\
$^{69}$ Institute of Space Science (ISS), Bucharest, Romania\\
$^{70}$ Institut f\"{u}r Kernphysik, Johann Wolfgang Goethe-Universit\"{a}t Frankfurt, Frankfurt, Germany\\
$^{71}$ Instituto de Ciencias Nucleares, Universidad Nacional Aut\'{o}noma de M\'{e}xico, Mexico City, Mexico\\
$^{72}$ Instituto de F\'{i}sica, Universidade Federal do Rio Grande do Sul (UFRGS), Porto Alegre, Brazil\\
$^{73}$ Instituto de F\'{\i}sica, Universidad Nacional Aut\'{o}noma de M\'{e}xico, Mexico City, Mexico\\
$^{74}$ iThemba LABS, National Research Foundation, Somerset West, South Africa\\
$^{75}$ Jeonbuk National University, Jeonju, Republic of Korea\\
$^{76}$ Johann-Wolfgang-Goethe Universit\"{a}t Frankfurt Institut f\"{u}r Informatik, Fachbereich Informatik und Mathematik, Frankfurt, Germany\\
$^{77}$ Joint Institute for Nuclear Research (JINR), Dubna, Russia\\
$^{78}$ Korea Institute of Science and Technology Information, Daejeon, Republic of Korea\\
$^{79}$ KTO Karatay University, Konya, Turkey\\
$^{80}$ Laboratoire de Physique des 2 Infinis, Ir\`{e}ne Joliot-Curie, Orsay, France\\
$^{81}$ Laboratoire de Physique Subatomique et de Cosmologie, Universit\'{e} Grenoble-Alpes, CNRS-IN2P3, Grenoble, France\\
$^{82}$ Lawrence Berkeley National Laboratory, Berkeley, California, United States\\
$^{83}$ Lund University Department of Physics, Division of Particle Physics, Lund, Sweden\\
$^{84}$ Moscow Institute for Physics and Technology, Moscow, Russia\\
$^{85}$ Nagasaki Institute of Applied Science, Nagasaki, Japan\\
$^{86}$ Nara Women{'}s University (NWU), Nara, Japan\\
$^{87}$ National and Kapodistrian University of Athens, School of Science, Department of Physics , Athens, Greece\\
$^{88}$ National Centre for Nuclear Research, Warsaw, Poland\\
$^{89}$ National Institute of Science Education and Research, Homi Bhabha National Institute, Jatni, India\\
$^{90}$ National Nuclear Research Center, Baku, Azerbaijan\\
$^{91}$ National Research Centre Kurchatov Institute, Moscow, Russia\\
$^{92}$ Niels Bohr Institute, University of Copenhagen, Copenhagen, Denmark\\
$^{93}$ Nikhef, National institute for subatomic physics, Amsterdam, Netherlands\\
$^{94}$ NRC Kurchatov Institute IHEP, Protvino, Russia\\
$^{95}$ NRC \guillemotleft Kurchatov\guillemotright  Institute - ITEP, Moscow, Russia\\
$^{96}$ NRNU Moscow Engineering Physics Institute, Moscow, Russia\\
$^{97}$ Nuclear Physics Group, STFC Daresbury Laboratory, Daresbury, United Kingdom\\
$^{98}$ Nuclear Physics Institute of the Czech Academy of Sciences, \v{R}e\v{z} u Prahy, Czech Republic\\
$^{99}$ Oak Ridge National Laboratory, Oak Ridge, Tennessee, United States\\
$^{100}$ Ohio State University, Columbus, Ohio, United States\\
$^{101}$ Petersburg Nuclear Physics Institute, Gatchina, Russia\\
$^{102}$ Physics department, Faculty of science, University of Zagreb, Zagreb, Croatia\\
$^{103}$ Physics Department, Panjab University, Chandigarh, India\\
$^{104}$ Physics Department, University of Jammu, Jammu, India\\
$^{105}$ Physics Department, University of Rajasthan, Jaipur, India\\
$^{106}$ Physikalisches Institut, Eberhard-Karls-Universit\"{a}t T\"{u}bingen, T\"{u}bingen, Germany\\
$^{107}$ Physikalisches Institut, Ruprecht-Karls-Universit\"{a}t Heidelberg, Heidelberg, Germany\\
$^{108}$ Physik Department, Technische Universit\"{a}t M\"{u}nchen, Munich, Germany\\
$^{109}$ PINSTECH, Islamabad, Pakistan\\
$^{110}$ Politecnico di Bari and Sezione INFN, Bari, Italy\\
$^{111}$ Research Division and ExtreMe Matter Institute EMMI, GSI Helmholtzzentrum f\"ur Schwerionenforschung GmbH, Darmstadt, Germany\\
$^{112}$ Russian Federal Nuclear Center (VNIIEF), Sarov, Russia\\
$^{113}$ Saha Institute of Nuclear Physics, Homi Bhabha National Institute, Kolkata, India\\
$^{114}$ School of Physics and Astronomy, University of Birmingham, Birmingham, United Kingdom\\
$^{115}$ Secci\'{o}n F\'{\i}sica, Departamento de Ciencias, Pontificia Universidad Cat\'{o}lica del Per\'{u}, Lima, Peru\\
$^{116}$ St. Petersburg State University, St. Petersburg, Russia\\
$^{117}$ Stefan Meyer Institut f\"{u}r Subatomare Physik (SMI), Vienna, Austria\\
$^{118}$ SUBATECH, IMT Atlantique, Universit\'{e} de Nantes, CNRS-IN2P3, Nantes, France\\
$^{119}$ Suranaree University of Technology, Nakhon Ratchasima, Thailand\\
$^{120}$ Technical University of Ko\v{s}ice, Ko\v{s}ice, Slovakia\\
$^{121}$ The Henryk Niewodniczanski Institute of Nuclear Physics, Polish Academy of Sciences, Cracow, Poland\\
$^{122}$ The University of Texas at Austin, Austin, Texas, United States\\
$^{123}$ Universidad Aut\'{o}noma de Sinaloa, Culiac\'{a}n, Mexico\\
$^{124}$ Universidade de S\~{a}o Paulo (USP), S\~{a}o Paulo, Brazil\\
$^{125}$ Universidade Estadual de Campinas (UNICAMP), Campinas, Brazil\\
$^{126}$ Universidade Federal do ABC, Santo Andre, Brazil\\
$^{127}$ University of Cape Town, Cape Town, South Africa\\
$^{128}$ University of Houston, Houston, Texas, United States\\
$^{129}$ University of Jyv\"{a}skyl\"{a}, Jyv\"{a}skyl\"{a}, Finland\\
$^{130}$ University of Kansas, Lawrence, Kansas, United States\\
$^{131}$ University of Liverpool, Liverpool, United Kingdom\\
$^{132}$ University of Science and Technology of China, Hefei, China\\
$^{133}$ University of South-Eastern Norway, Tonsberg, Norway\\
$^{134}$ University of Tennessee, Knoxville, Tennessee, United States\\
$^{135}$ University of the Witwatersrand, Johannesburg, South Africa\\
$^{136}$ University of Tokyo, Tokyo, Japan\\
$^{137}$ University of Tsukuba, Tsukuba, Japan\\
$^{138}$ Universit\'{e} Clermont Auvergne, CNRS/IN2P3, LPC, Clermont-Ferrand, France\\
$^{139}$ Universit\'{e} de Lyon, CNRS/IN2P3, Institut de Physique des 2 Infinis de Lyon, Lyon, France\\
$^{140}$ Universit\'{e} de Strasbourg, CNRS, IPHC UMR 7178, F-67000 Strasbourg, France, Strasbourg, France\\
$^{141}$ Universit\'{e} Paris-Saclay Centre d'Etudes de Saclay (CEA), IRFU, D\'{e}partment de Physique Nucl\'{e}aire (DPhN), Saclay, France\\
$^{142}$ Universit\`{a} degli Studi di Foggia, Foggia, Italy\\
$^{143}$ Universit\`{a} di Brescia, Brescia, Italy\\
$^{144}$ Variable Energy Cyclotron Centre, Homi Bhabha National Institute, Kolkata, India\\
$^{145}$ Warsaw University of Technology, Warsaw, Poland\\
$^{146}$ Wayne State University, Detroit, Michigan, United States\\
$^{147}$ Westf\"{a}lische Wilhelms-Universit\"{a}t M\"{u}nster, Institut f\"{u}r Kernphysik, M\"{u}nster, Germany\\
$^{148}$ Wigner Research Centre for Physics, Budapest, Hungary\\
$^{149}$ Yale University, New Haven, Connecticut, United States\\
$^{150}$ Yonsei University, Seoul, Republic of Korea\\

\end{flushleft} 
  
\end{document}